\documentclass[12pt]{iopart}
\usepackage{iopams}  
\usepackage{revsymb}
\usepackage{graphicx}
\usepackage{lscape}
\begin{document}

\newcommand{\apj}{{Astrophys.\ J. }}
\newcommand{\apjs}{{Astrophys.\ J.\ Supp. }}
\newcommand{\apjl}{{Astrophys.\ J.\ Lett. }}
\newcommand{\aj}{{Astron.\ J. }}
\newcommand{\prl}{{Phys.\ Rev.\ Lett. }}
\newcommand{\prd}{{Phys.\ Rev.\ D }}
\newcommand{\mnras}{{Mon.\ Not.\ R.\ Astron.\ Soc. }}
\newcommand{\araa}{{ARA\&A }}
\newcommand{\aap}{{Astron.\ \& Astrophy. }}
\newcommand{\nat}{{Nature }}
\newcommand{\cqg}{{Class.\ Quantum Grav.\ }}

\def\HS{{\mathfrak H}_3}
\def\bfis{\hbox{\scriptsize\rm i}}
\def\bfi{\hbox{\rm i}}
\def\bfj{\hbox{\rm j}}
\def\3{{\ss}}

\hfill astro-ph/0504656

\title[CMB Anisotropy of Spherical Spaces]
{CMB Anisotropy of Spherical Spaces}

\author{Ralf Aurich$^1$, Sven Lustig$^1$, Frank Steiner$^{1,2}$}

\address{$^1$Abteilung Theoretische Physik, Universit\"at Ulm,\\
Albert-Einstein-Allee 11, D-89069 Ulm, Germany}

\vspace*{3pt}
\address{$^2$Theory Division, Physics Department,\\
         CERN, CH-1211 Geneva 23, Switzerland}

\begin{abstract}
The first-year WMAP data taken at their face value hint that the Universe
might be slightly positively curved and therefore necessarily finite,
since all spherical (Clifford-Klein) space forms
${\cal M}^3 = {\cal S}^3/\Gamma$,
given by the quotient of ${\cal S}^3$ by a group $\Gamma$ of
covering transformations, possess this property.
We examine the anisotropy of the cosmic microwave background (CMB)
for all typical groups $\Gamma$ corresponding to homogeneous universes.
The CMB angular power spectrum and the temperature correlation function
are computed for the homogeneous spaces as a function of the
total energy density parameter $\Omega_{\hbox{\scriptsize tot}}$
in the large range $[1.01, 1.20]$ and are compared with the WMAP data.
We find that out of the infinitely many homogeneous spaces only the
three corresponding to the binary dihedral group $T^\star$,
the binary octahedral group $O^\star$,
and the binary icosahedral group $I^\star$ are in agreement
with the WMAP observations.
Furthermore, if $\Omega_{\hbox{\scriptsize tot}}$ is restricted
to the interval $[1.00, 1.04]$,
the space described by $T^\star$ is excluded since it requires a value
of $\Omega_{\hbox{\scriptsize tot}}$
which is probably too large being in the range $[1.06, 1.07]$.
We thus conclude that there remain only the two homogeneous
spherical spaces ${\cal S}^3/O^\star$ and
${\cal S}^3/I^\star$ with $\Omega_{\hbox{\scriptsize tot}}$ of about
1.038 and 1.018, respectively, as possible topologies for our Universe.
\end{abstract}

\pacs{98.80.-k, 98.70.Vc, 98.80.Es}

%\submitto{\CQG}

% \maketitle

%%%%%%%%%%%%%%%%%%%%%%%%%%%%%%%%%%%%%%%%%%%%%%%%%%%%%%%%%%%%%%%%%%%%%%%%%%%%
%%%%%%%%%%%%%%%%%%%%%%%%%%%%%%%%%%%%%%%%%%%%%%%%%%%%%%%%%%%%%%%%%%%%%%%%%%%%

\section{Introduction}

At present, all data are consistent with,
and in fact strongly support, the standard big-bang model
in which the time evolution of the Universe is described by the
Friedmann-Lema\^{\i}tre-Robertson-Walker metric.
Accordingly, the Universe possesses the space-time structure
$\mathbb{R}\times{\cal M}$
where $\mathbb{R}$ describes the ``space'' of cosmic time,
and ${\cal M}$ the three-dimensional comoving space section of constant
curvature $K=+1, 0$ and $-1$.
The Einstein field equations respectively
the Friedmann equations for the cosmic scale factor do not fix
the curvature a priori.
Instead, the curvature parameter $K$ has to be inferred from a
determination of the total energy density
$\varepsilon_{\hbox{\scriptsize tot}}$
of the Universe via the relation $(c=1)\,$
$K = H_0^2 a_0^2(\Omega_{\hbox{\scriptsize tot}}-1)$,
$\Omega_{\hbox{\scriptsize tot}}:=\varepsilon_{\hbox{\scriptsize tot}}/
\varepsilon_{\hbox{\scriptsize crit}}$,
where $\varepsilon_{\hbox{\scriptsize crit}} := \frac{3H_0^2}{8\pi G}$
denotes the critical energy density,
$a_0$ the cosmic scale factor, and $H_0$ the Hubble constant
(all quantities at the present epoch).
Furthermore, it is a mathematical fact,
although not always appreciated
(see, however, the remark on early works below),
that fixing the curvature $K$ does not determine uniquely
the global geometry of ${\cal M}$,
i.\,e.\ the topology and thus the shape of the Universe.
Only if it is {\it assumed} that the Universe is simply-connected,
the possible homogeneous 3-spaces ${\cal M}$ of constant curvature $K$
are given by the 3-sphere ${\cal S}^3 (K=+1)$,
Euclidean 3-space ${\cal E}^3 (K=0)$,
or hyperbolic 3-space ${\cal H}^3 (K=-1)$.
In this case, the Universe is finite for positive curvature
$(\Omega_{\hbox{\scriptsize tot}}>1)$
and infinite for vanishing $(\Omega_{\hbox{\scriptsize tot}}=1)$
or negative curvature $(\Omega_{\hbox{\scriptsize tot}}<1)$.
However, most 3-spaces ${\cal M}$ of constant curvature are
multi-connected and are given by the quotient of
${\cal S}^3$, ${\cal E}^3$, or ${\cal H}^3$ by a group $\Gamma$
of covering transformations,
i.\,e.\ ${\cal M}={\cal S}^3/\Gamma$, ${\cal E}^3/\Gamma$, or
${\cal H}^3/\Gamma$.
In this case, the Universe is again finite for positive curvature,
but can be finite too if it is flat or negatively curved.

Here, we would like to remark that the question whether the space of the
Universe is finite and possibly multi-connected has been discussed
during the last century by several cosmologists,
e.\,g.\ by Schwarzschild \cite{Schwarzschild_1900},
Einstein \cite{Einstein_1917},
Friedmann \cite{Friedmann_1922,Friedmann_1924},
Lema\^{\i}tre \cite{Lemaitre_1958},
Heckmann and Sch\"ucking \cite{Heckmann_Schuecking_1959a},
and Ellis \cite{Ellis_1971}, to mention only a few.

The concordance model of cosmology ($\Lambda$CDM model)
assumes a flat Universe with the topology of ${\cal E}^3$
with a positive cosmological constant $\Lambda$,
i.\,e.\ $\Omega_\Lambda := \frac{\Lambda}{3H_0^2} =
1 -\Omega_{\hbox{\scriptsize mat}}-\Omega_{\hbox{\scriptsize rad}}$
with $\Omega_{\hbox{\scriptsize mat}}=\Omega_{\hbox{\scriptsize bar}}
+ \Omega_{\hbox{\scriptsize cdm}}$,
where the various $\Omega$-parameters denote the present value
of the baryonic (bar), cold dark matter (cdm), matter (mat)
and radiation (rad) energy densities in units of
$\varepsilon_{\hbox{\scriptsize crit}}$.
Three variants of the concordance model have been presented
by the WMAP team \cite{Spergel_et_al_2003}
providing a good overall fit to the temperature fluctuations
$\delta T$ of the cosmic microwave background radiation (CMB)
on small and medium scales, but there remains a strange
discrepancy at large scales as first observed by COBE
\cite{Hinshaw_et_al_1996}
and later substantiated by WMAP \cite{Bennett_et_al_2003}.

The suppression of the CMB anisotropy at large scales
respectively low multipoles can be explained
if the Universe is finite.
Recent analyses concerning the suppression
at low multipoles in the WMAP data can be found in
\cite{Luminet_Weeks_Riazuelo_Lehoucq_Uzan_2003,%
Aurich_Lustig_Steiner_Then_2004a,Aurich_Lustig_Steiner_Then_2004b,%
Aurich_Lustig_Steiner_2004c,Gundermann_2005}.
As discussed before, a finite Universe is naturally obtained,
if the total energy density exceeds the critical value one,
i.\,e.\ $\Omega_{\hbox{\scriptsize tot}}>1$.
Interestingly enough, the WMAP team reported \cite{Bennett_et_al_2003}
$\Omega_{\hbox{\scriptsize tot}}=1.02\pm0.02$
together with $\Omega_{\hbox{\scriptsize bar}}=0.044\pm0.004$,
$\Omega_{\hbox{\scriptsize mat}}=0.27\pm0.04$,
and $h=0.71^{+0.04}_{-0.03}$ for the present day reduced Hubble constant
(the errors give the $1\sigma$-deviation uncertainties only).
Taking at their face value, these parameters hint to a
positively curved Universe possessing the geometry of ${\cal S}^3$
or of one of the spatial space forms ${\cal S}^3/\Gamma$.
(One should keep in mind, however, that the WMAP values depend on certain
priors, and, furthermore, include the $1\sigma$-errors only.
Thus it would be too early to conclude that the data definitively
exclude a negatively curved Universe.
In fact, we have recently shown
\cite{Aurich_Lustig_Steiner_Then_2004a,Aurich_Lustig_Steiner_Then_2004b}
that the non-compact, but finite hyperbolic Picard universe
describes well the CMB anisotropy and the observed suppression
of power at large scales.)

In this paper, we present a systematic comparison of the predictions
with the CMB anisotropy for universes possessing homogeneous
spherical topology.
This comparison is made possible for two reasons.
First of all, the spherical spaces were classified already by 1932
\cite{Threlfall_Seifert_1930,Threlfall_Seifert_1932}
and thus their mathematical structure is known.
(This is in contrast to the case of hyperbolic manifolds
which are not yet completely classified; even the manifold with the
smallest volume is not yet known.)
Second, due to an efficient numerical algorithm described in
our recent paper \cite{Aurich_Lustig_Steiner_2004c},
we are able to take in the Sachs-Wolfe formula a large number of
vibrational modes into account
and thus to predict sufficiently many CMB multipoles for the
various  spherical spaces which in turn allow a detailed
comparison with the WMAP data.
Since the CMB spectrum depends sensitively on the curvature radius,
the comparison is performed as a function of
$\Omega_{\hbox{\scriptsize tot}}$ in the large interval $[1.01,1.20]$
in order to determine for a given spherical space form the
best-fitting value of the total energy density -
under the condition, of course, that the space under consideration
is able to describe the data at all.

Recently, Luminet et al.\ \cite{Luminet_Weeks_Riazuelo_Lehoucq_Uzan_2003}
proposed the Poincar\'e dodecahedron, which is one of the well-known
spherical space forms
(see section \ref{Section_spherical_space_form} for details),
as a model for the geometry of the Universe.
In their preliminary study involving only the three lowest multipoles
$(l=2,3,4)$ they found, indeed,
for $\Omega_{\hbox{\scriptsize tot}} = 1.013$ a strong suppression
of the CMB power at $l=2$ and a weak suppression at $l=3$ in agreement
with the WMAP data.
However, in \cite{Luminet_Weeks_Riazuelo_Lehoucq_Uzan_2003}
only the first three  vibrational modes of the dodecahedral space
with wave number $\beta=13, 21$ and 25
(comprising in total 59 eigenfunctions) have been used,
and there thus remained the question about how this extremely low
wave number cut-off affects the predictions of the multipoles,
since experience shows that increasing the cut-off usually enhances
the integrated Sachs-Wolfe and Doppler contributions.

In our recent paper \cite{Aurich_Lustig_Steiner_2004c}
we presented a thorough discussion of the CMB anisotropy for the
dodecahedral space topology using the first 10521 eigenfunctions
corresponding to the large wave number cut-off $\beta=155$.
The contributions of higher wave numbers up to $\beta=1501$ were
taken into account with respect to their mean behaviour.
Taking within the tight-coupling approximation not only the
ordinary, but also the integrated Sachs-Wolfe and also the
Doppler contribution into account, we were able to predict
sufficiently many multipole moments
such that a detailed comparison  of the dodecahedral space model
with the WMAP data could be performed.
We found that the temperature correlation function  for the
dodecahedral universe possesses very weak correlations at large
scales in nice agreement with the WMAP data for
$\Omega_{\hbox{\scriptsize tot}}$ in the range $1.016\dots1.020$.
There thus arises the interesting question whether the
dodecahedral space is the only spherical space form able to
describe the CMB data.
In \cite{Gundermann_2005} the CMB anisotropy is also studied
for the dodecahedron, the binary octahedral group, and
the binary tetrahedral group.

The main result of the present paper is that while almost all homogeneous
spherical spaces have to be excluded as possible geometries for the Universe,
there is one particular space form, defined by the binary octahedral group,
which agrees for $\Omega_{\hbox{\scriptsize tot}} \simeq 1.038$
with the WMAP data even better than the dodecahedron.
We observe that the best-fitting values obtained for
$\Omega_{\hbox{\scriptsize tot}}$ are different for the
binary octahedral space and the dodecahedron,
but both values lie well within the $1\sigma$-band
determined by WMAP.
It remains to be seen whether future data will enable us to
definitely eliminate one of the two space forms in favour of the
other as describing the true topology of the Universe.

Our paper is organised as follows.
In section \ref{Section_spherical_space_form},
we summarize the main properties of the existing homogeneous
spherical space forms and of their vibrational modes.
Our main results are presented in section \ref{Section_statistic}
which contains a detailed comparison with the WMAP data of the
CMB angular power spectrum for the various types of spherical spaces.
In addition to the power spectrum,
we study also the so-called $S(\rho)$ statistic \cite{Bennett_et_al_2003}
which measures the suppression at large angular scales directly
in terms of the temperature correlation function.
Section \ref{Section_Conclusion} contains our conclusions.

%%%%%%%%%%%%%%%%%%%%%%%%%%%%%%%%%%%%%%%%%%%%%%%%%%%%%%%%%%%%%%%
\section{The spherical space forms and their vibrational modes}
%%%%%%%%%%%%%%%%%%%%%%%%%%%%%%%%%%%%%%%%%%%%%%%%%%%%%%%%%%%%%%%
\label{Section_spherical_space_form}

In section 2 of \cite{Aurich_Lustig_Steiner_2004c} we have already
described the three-dimensional spaces ${\cal M}$ of constant
positive curvature $K=1$, and therefore we refer the reader to
this paper for details.
The spherical spaces were classified by 1932
\cite{Threlfall_Seifert_1930,Threlfall_Seifert_1932}
and are given by the quotient ${\cal M} = {\cal S}^3/\Gamma$
of the three-sphere ${\cal S}^3$ under the action of a discrete
fixed-point free subgroup $\Gamma \subset \hbox{SO}(4)$
of the isometries of ${\cal S}^3$.
All these manifolds are compact possessing the volume
$V({\cal S}^3/\Gamma) = V({\cal S}^3)/N$,
where $N$ is the order of the group $\Gamma$,
and are, apart from the universal covering space ${\cal S}^3$ with
volume $V({\cal S}^3)=2\pi^2$, multi-connected.
To define the discrete fixed-point free subgroups
$\Gamma \subset \hbox{SO}(4)$ of isometries of ${\cal S}^3$,
one makes use of the fact that the unit 3-sphere ${\cal S}^3$
can be identified with the multiplicative group of unit quaternions $\{q\}$.
The latter are defined by $q:=w+x\bfi+y\bfj+z\bfi\bfj$,
$(w,x,y,z)\in \mathbb{R}^4$, having unit norm, $|q|^2 = w^2+x^2+y^2+z^2=1$.
Here, the 4 basic quaternions $\{1,\bfi,\bfj,\bfi\bfj\}$ satisfy
the multiplication rules $\bfi^2=\bfj^2=-1$ and $\bfi\bfj=-\bfj\bfi$
plus the property that $\bfi$ and $\bfj$ commute with every real number.
The distance $d(q_1,q_2)$ between two points $q_1$ and $q_2$ on ${\cal S}^3$
is given by $\cos d(q_1,q_2) = w_1w_2+x_1x_2+y_1y_2+z_1z_2$.

The group $\hbox{SO}(4)$ is isomorphic to
${\cal S}^3 \times {\cal S}^3 / \{\pm(1,1)\}$,
the two factors corresponding to the left and right group actions.
In this paper, we are only interested in homogeneous manifolds
${\cal M} = {\cal S}^3/\Gamma$, in which case the group $\Gamma$
contains only right-handed Clifford translations $\gamma\in\Gamma$
that act on an arbitrary unit quaternion $q\in{\cal S}^3$ by
left-multiplication, $q \to \gamma q$, and translate all points
$q_1,q_2\in {\cal S}^3$ by the same distance $\chi$,
i.\,e.\ $d(q_1,\gamma q_1)=d(q_2,\gamma q_2)=\chi$.
The right-handed Clifford translations act as right-handed
cork screw fixed-point free rotations of ${\cal S}^3$.
The following groups lead to homogeneous manifolds
${\cal M} = {\cal S}^3/\Gamma$
\cite{Threlfall_Seifert_1930,Threlfall_Seifert_1932,Wolf_1974,Thurston_1997}:
\begin{itemize}
\item The cyclic groups $Z_m$ of order $m$ $(m\ge 1)$.
\item The binary dihedral groups $D_{4m}^\star$ of order $4m$ $(m\ge 2)$.
\item The binary tetrahedral group $T^\star$ of order $24$.
\item The binary octahedral group $O^\star$ of order $48$.
\item The binary icosahedral group $I^\star$ of order $120$.
\end{itemize}
In table \ref{Tab:Generators} we give the right-handed Clifford translations
which generate the above groups $\Gamma$.

\begin{table}
\hspace*{2cm}\begin{tabular}{|c|c|c|}
\hline
$\Gamma$ & $\gamma_1$ & $\gamma_2$ \\
\hline
$Z_m$ & $\cos\left(\frac{2\pi}m\right) +
    \bfi \, \sin\left(\frac{2\pi}m\right)$ & $-$ \\
\hline
$D_{4m}^\star $ & $\cos\left(\frac{2\pi}m\right) +
    \bfi\bfj \, \sin\left(\frac{2\pi}m\right)$ & $\bfi$  \\
\hline
$T^\star$ & $\bfj$ &
$\frac 12 + \frac 12 \bfi + \frac 12 \bfj + \frac 12 \bfi\bfj$ \\
\hline
$O^\star$ & $\frac 1{\sqrt 2} + \frac 1{\sqrt 2}\bfi$ &
$\frac 12 + \frac 12 \bfi + \frac 12 \bfj + \frac 12 \bfi\bfj$ \\
\hline
$I^\star$ & $\bfj$ & $\frac{\sigma}2 + \frac 1{2\sigma}\bfi + \frac 12\bfj$ \\
\hline
\end{tabular}
\caption{\label{Tab:Generators}
The generators $\gamma_1$ and $\gamma_2$ for the groups $\Gamma$
$(\sigma = (1 + \sqrt 5)/2)$.
}
\end{table}

The vibrations on the homogeneous spherical spaces
${\cal M} = {\cal S}^3/\Gamma$ are determined by the regular solutions
of the Helmholtz equation
\begin{equation}
\label{Eq:Helmholtz}
(\Delta + E_\beta^{\cal M}) \, \psi_\beta^{{\cal M},i}(q) \; = \; 0
\hspace{10pt} , \hspace{10pt}
q \in {\cal M} \; \; ,
\end{equation}
satisfying the fundamental periodicity conditions
\begin{equation}
\label{Eq:periodicity_condition}
\psi_\beta^{{\cal M},i}(\gamma_k q) \; = \; \psi_\beta^{{\cal M},i}(q)
\hspace{10pt} , \hspace{10pt}
\forall q \in {\cal M} \; \; , \; \; \forall \gamma_k \in \Gamma
\hspace{10pt} .
\end{equation}
Here $\Delta$ denotes the Laplace-Beltrami operator on ${\cal S}^3$.
The eigenfunctions on ${\cal M}$ satisfy the orthonormality relation
\begin{equation}
\label{Eq:orthonormality}
\int_{\cal M} d\mu \;
\psi_\beta^{{\cal M},i}(q) \; \psi_{\beta'}^{{\cal M},i'}(q) \; = \;
\frac 1N \, \int_{{\cal S}^3} d\mu \;
\psi_\beta^{{\cal M},i}(q) \; \psi_{\beta'}^{{\cal M},i'}(q) \; = \;
\delta_{\beta\beta'} \, \delta{ii'}
\hspace{10pt} .
\end{equation}
The spectrum on ${\cal M}$ is discrete, and the eigenvalues can be
expressed in terms of the wave number $\beta\in\mathbb{N}$ as
$E_\beta = \beta^2-1$ and are independent
of the degeneracy index $i=1,\dots,r^{\cal M}(\beta)$,
where $r^{\cal M}(\beta)$ denotes the multiplicity of the mode $\beta$.
It should be noted that for a given manifold ${\cal M}$
the wave numbers $\beta$ do not take all values in $\mathbb{N}$.
The allowed $\beta$ values together with their multiplicities
$r^{\cal M}(\beta)$ are explicitly known \cite{Ikeda_1995,Weeks_2005},
see Table \ref{Tab:Spectrum}.

The eigenfunctions $\psi_\beta^{{\cal M},i}(q)$ on ${\cal M}$
can be expanded into the eigenfunctions
$\psi_{\beta lm}^{{\cal S}^3}(q)$ on ${\cal S}^3$
\begin{equation}
\label{Eq:expansion_in_S3}
\psi_\beta^{{\cal M},i}(q) \; = \;
\sum_{l=0}^{\beta-1} \sum_{m=-l}^l \xi_{\beta lm}^i({\cal M}) \,
\psi_{\beta lm}^{{\cal S}^3}(q)
\hspace{10pt} .
\end{equation}
Since the eigenfunctions on ${\cal S}^3$ are explicitly known
\cite{Schroedinger_1938,Schroedinger_1939,Schroedinger_1940a,%
Schroedinger_1940b,Harrison_1967,Abbott_Schaefer_1986},
it remains to determine the expansion coefficients
$\xi_{\beta lm}^i({\cal M})$
which satisfy as a consequence of eq.\,(\ref{Eq:orthonormality})
the normalization condition
\begin{equation}
\label{Eq:normalization_condition}
\sum_{l=0}^{\beta-1} \sum_{m=-l}^l \left(\xi_{\beta lm}^i({\cal M})\right)^2
\; = \; N
\hspace{10pt} .
\end{equation}
In \cite{Aurich_Lustig_Steiner_2004c} we have described our
numerical algorithm to compute the coefficients $\xi_{\beta lm}^i({\cal M})$.
It uses a collocation method by imposing the periodicity condition
(\ref{Eq:periodicity_condition}).
Using this method, we have computed in \cite{Aurich_Lustig_Steiner_2004c}
the expansion coefficients for $\Gamma=I^\star$,
i.\,e.\ for the Poincar\'e dodecahedral space
${\cal D} = {\cal S}^3/I^\star$,
for $\beta\leq 155$ comprising the first 10521 eigenfunctions.
In addition, we have computed in \cite{Aurich_Lustig_Steiner_2004c}
the coefficients $\xi_{\beta lm}^i({\cal M})$ for $\beta\leq 33$
for some cyclic groups, some binary dihedral groups,
the binary tetrahedral group, and the binary octahedral group.
Based on these numerical results, we stated for homogeneous space forms
the following relation as a conjecture $(0\leq l \leq \beta-1)$
\begin{equation}
\label{Eq:Conjecture}
\frac 1{2l+1} \sum_{m=-l}^l  \sum_{i=1}^{r^{\cal M}(\beta)}
\left( \xi_{\beta l m}^i({\cal M}) \right)^2 \; = \;
N \, \frac{r^{\cal M}(\beta)}{r^{{\cal S}^3}(\beta)}
\hspace{10pt} ,
\end{equation}
where $r^{{\cal S}^3}(\beta)=\beta^2$ denotes the multiplicity of the
vibrational modes on ${\cal S}^3$.
The relation (\ref{Eq:Conjecture}) has been found to hold within a numerical
accuracy of 13 digits.
However, for the inhomogeneous lens spaces
\cite{Gausmann_Lehoucq_Luminet_Uzan_Weeks_2001}
L(12,5) and L(72,17) we have found that the relation (\ref{Eq:Conjecture})
does not hold.
We thus concluded in \cite{Aurich_Lustig_Steiner_2004c}
that the relation (\ref{Eq:Conjecture}) is only valid for
homogeneous 3-spaces.
Recently, Gundermann \cite{Gundermann_2005} provided a proof of our
conjecture (\ref{Eq:Conjecture}).

In the next section, we shall use the eigenfunctions
(\ref{Eq:expansion_in_S3}) to calculate CMB sky maps
and the variance of the CMB anisotropy
and relation (\ref{Eq:Conjecture}) to calculate the
mean value of the CMB anisotropy for a variety of spherical spaces.

%%%%%%%%%%%%%%%%%%%%%%%%%%%%%%%%%%%%%%%%%%%%%%%%%%%%%%%%%%%%%%%%%%
\section{The angular power spectrum $\delta T_l^2$ and the
         correlation function $C(\vartheta)$ for spherical spaces}
%%%%%%%%%%%%%%%%%%%%%%%%%%%%%%%%%%%%%%%%%%%%%%%%%%%%%%%%%%%%%%%%%%
\label{Section_statistic}

The relative temperature fluctuations $\frac{\delta T(\hat n)}T$
of the CMB are caused by several effects
which we shall compute within the tight-coupling approximation
along the lines described in detail in Section 2 of
\cite{Aurich_Lustig_Steiner_Then_2004a}.
The dominant contribution at large scales is given by the
ordinary Sachs-Wolfe (SW) effect which is a combination of
the gravitational potential $\Phi(\eta,\tau,\theta,\phi)$
at the surface of last scattering (SLS), and the intrinsic temperature
fluctuation $\frac 14\delta_\gamma(\eta,\tau,\theta,\phi)$
due to the imposed entropic initial conditions,
where $\delta_\gamma$ denotes the relative perturbation in
the radiation component.
(Here $\eta$ denotes the conformal time and $(\tau(\eta),\theta,\phi)$
the spherical coordinates of the photon path in the direction
$\hat n = \hat n(\theta,\phi)$,
where we assume that the observer is at the origin of the coordinate system,
i.\,e.\ at $(\tau_{\hbox{\scriptsize obs}},\theta_{\hbox{\scriptsize obs}},
\phi_{\hbox{\scriptsize obs}}) = (0,0,0)$.)

The gravitational potential $\Phi$ is identified with the (scalar)
perturbation of the Friedmann-Lema\^{\i}tre-Robertson-Walker metric
which for an energy-momentum tensor $T_{\mu\nu}$ with
$T_{ij}=0$ for $i\neq j$ $(i,j=1,2,3)$ can in conformal Newtonian gauge
be written as \cite{Bardeen_1980}
\begin{equation}
\label{Eq:Metric}
ds^2 \; = \; a^2(\eta) \, \left[ \, (1+2\Phi) d\eta^2 -
(1-2\Phi) |d\vec x|^2 \, \right]
\hspace{10pt} .
\end{equation}
Here $a(\eta)$ denotes the cosmic scale factor as a function of
conformal time $\eta$ and $|d\vec x|^2$ the line element on ${\cal S}^3$
\begin{equation}
\label{Eq:Metric_on_S3}
|d\vec x|^2 \; = \; d\tau^2 \, + \,
\sin^2\tau \; \big( d\theta^2 + \sin^2\theta \, d\phi^2 \big)
\end{equation}
with $0\leq \tau \leq \pi$, $0\leq \theta \leq \pi$, $0\leq \phi \leq 2\pi$. 

The ordinary Sachs-Wolfe (SW) contribution to the temperature fluctuation
is given by
\begin{equation}
\label{Eq:NSW}
\frac{\delta T^{\hbox{\scriptsize SW}}(\hat n)}T \; = \;
\Phi(\eta_{\hbox{\scriptsize SLS}},\tau_{\hbox{\scriptsize SLS}},\theta,\phi)
\, + \,
\frac 14 \delta_\gamma(\eta_{\hbox{\scriptsize SLS}},
\tau_{\hbox{\scriptsize SLS}},\theta,\phi)
\end{equation}
with $\tau_{\hbox{\scriptsize SLS}} := \eta_0 - \eta_{\hbox{\scriptsize SLS}}$,
where $\eta_0$ and $\eta_{\hbox{\scriptsize SLS}}$ denote the conformal time
at the present epoch and at the time of recombination corresponding
to a redshift $z_{\hbox{\scriptsize SLS}}=1089$, respectively.
For a given spherical space ${\cal M}$, the metric perturbation can be
written as an expansion in the eigenfunctions on ${\cal M}$
\begin{equation}
\label{Eq:grav_potential}
\Phi(\eta,\tau,\theta,\phi) \; = \;
{\sum_{\beta\ge 3}} '\; \sum_{i=1}^{r^{\cal M}(\beta)}
\Phi_\beta^i(\eta) \, \Psi_\beta^{{\cal M},i}(\tau,\theta,\phi)
\hspace{10pt} ,
\end{equation}
where in the mode summation only the modes with $\beta\ge 3$ have been
taken into account (if they exist),
since the wave numbers $\beta=1,2$ correspond to modes which are pure
gauge terms \cite{Bardeen_1980}.
The prime in the summation over the modes $\beta$ indicates
that the spectrum of a given manifold ${\cal M}$ does not
contain all $\beta\in\mathbb{N}$, see Table \ref{Tab:Spectrum}.
The functions $\Phi_\beta^i(\eta)$ determine the time evolution and
will be factorized
$\Phi_\beta^i(\eta) = \Phi_\beta^i(0) \, g_\beta(\eta)$
with $g_\beta(0)=1$.
The functions $g_\beta(\eta)$ do not depend on the degeneracy index $i$,
since the associated differential equation depends only on
the eigenvalue $E_\beta^{\cal M}$ which is independent of $i$.
The initial values $\Phi_\beta^i(0)$ are the primordial
fluctuation amplitudes and are assumed to be Gaussian random variables with
zero expectation value and covariance
\begin{equation}
\label{Eq:covariance}
\left< \Phi_\beta^i(0)\, \Phi_{\beta'}^{i'}(0) \right> \; = \;
\delta_{\beta\beta'} \, \delta_{i i'} \, P_\Phi(\beta)
\hspace{10pt} .
\end{equation}
Here $P_\Phi(\beta)$ denotes the primordial power spectrum
that determines the weight by which the primordial modes $\beta$
are excited, on average.
The average $\left<\dots\right>$ in (\ref{Eq:covariance}) denotes an
ensemble average over the primordial perturbations
which are supposed to arise from quantum fluctuations,
by which the Universe is ``created''.
In the following, we shall assume that the primordial power spectrum
is in good approximation described by the scale-invariant
Harrison-Zel'dovich spectrum
\begin{equation}
\label{Eq:Harrison_Zeldovich}
P_\Phi(\beta) \; = \; \frac \alpha{\beta(\beta^2-1)}
\hspace{10pt} .
\end{equation}
Here $\alpha$ is a normalization factor
which will be determined from the CMB data.

The {\it temperature fluctuations} $\delta T(\hat n)$
of the microwave sky can be expanded into real spherical harmonics
$\tilde Y_{lm}(\hat n)$ on ${\cal S}^2$,
\begin{equation}
\label{Eq:delta_T_expansion}
\delta T(\hat n) \; := \;
\sum_{l=2}^\infty \sum_{m=-l}^l \, a_{lm} \, \tilde Y_{lm}(\hat n)
\hspace{10pt} ,
\end{equation}
where the monopole and dipole terms, $l=0,1$, are not included
in the sum (\ref{Eq:delta_T_expansion}).
From the real expansion coefficients $a_{lm}$ one forms
the {\it multipole moments}
\begin{equation}
\label{Eq:C_l}
C_l \; := \;
\frac 1{2l+1} \, \Big< \sum_{m=-l}^l \, \big( a_{lm} \big)^2 \, \Big>
\end{equation}
and the {\it angular power spectrum}
\begin{equation}
\label{Eq:angular_power_spectrum}
\delta T_l^2 \; := \; \frac{l(l+1)}{2\pi} \, C_l
\hspace{10pt} .
\end{equation}
The average $\left< \dots \right>$ in (\ref{Eq:C_l}) denotes an
ensemble average over the primordial perturbations as in
eq.(\ref{Eq:covariance}),
respectively an ensemble average over the universal observers.

Inserting the expansions (\ref{Eq:grav_potential}) and
(\ref{Eq:expansion_in_S3}) into the approximation
$\frac{\delta T^{\hbox{\scriptsize SW}}(\hat n)}T \simeq
\frac 13 \Phi(\eta_{\hbox{\scriptsize SLS}},\tau_{\hbox{\scriptsize SLS}},
\theta,\phi)$
to the Sachs-Wolfe formula (\ref{Eq:NSW})
and using the explicit expression for the eigenfunctions
$\psi_{\beta lm}^{{\cal S}^3}$ on ${\cal S}^3$,
\begin{equation}
\label{Eq:modes_S3}
\psi_{\beta lm}^{{\cal S}^3}(\vec x\,) \; = \;
R_{\beta l}(\tau) \, \tilde Y_{lm}(\theta,\phi)
\hspace{10pt} ,
\end{equation}
one arrives \cite{Aurich_Lustig_Steiner_2004c}
with the help of relation (\ref{Eq:Conjecture}) at the following
expression for the ordinary Sachs-Wolfe contribution to the multipole
moments for a given spherical space ${\cal M} = {\cal S}^3/\Gamma$
\begin{equation}
\label{Eq:Cl_NSW}
C_l^{\hbox{\scriptsize SW}}({\cal M}) \; = \;
\frac{N}{9} {\sum_{\beta > l}}' \;
\frac{r^{\cal M}(\beta)}{\beta^2} \, P_\Phi(\beta) \,
g_\beta^2(\eta_{\hbox{\scriptsize SLS}}) \,
R_{\beta l}^2(\tau_{\hbox{\scriptsize SLS}})
\hspace{10pt} .
\end{equation}
Here $R_{\beta l}(\tau)$ denote the
``radial functions'' on ${\cal S}^3$
which can be expressed in terms of Gegenbauer polynomials,
see eq.(10) of \cite{Aurich_Lustig_Steiner_2004c}.
The expression (\ref{Eq:Cl_NSW}) shows in a transparent way,
why the lowest multipoles are in general suppressed for the
multi-connected spherical spaces,
the more the more wave numbers $\beta$ are missing in
the vibrational spectrum.
Let us consider, for example, the quadrupole moment, $l=2$.
Then the summation over modes in (\ref{Eq:Cl_NSW}) runs for the
simply-connected manifold ${\cal S}^3$ over all natural numbers
with $\beta\ge 3$.
In contrast, for the Poincar\'e dodecahedral space ${\cal D}$
there is a large gap between $\beta=3$ and 12, since the lowest
contributing mode occurs only at $\beta=13$, and thus the missing modes
lead to a suppression.
The suppression on ${\cal D}$ gets even stronger,
because there exist only the three modes $\beta=21,25$ and 31
in the wave number interval $13<\beta<33$
(see table \ref{Tab:Spectrum}).

Now, we would like to discuss this suppression mechanism in more detail.
Table \ref{Tab:Spectrum} shows that the allowed wave numbers $\beta$
are for all homogeneous spherical spaces given by odd natural numbers,
except for the homogeneous lens spaces ${\cal S}^3/Z_m$ with
$m$ odd $\ge 1$ for which $\beta$ runs through all natural numbers
with $\beta \ge m+1$, in addition to the lowest odd wave numbers
between 1 and $m$.
While the $\beta$ spectrum consists for the homogeneous lens spaces
$Z_m$ with $m$ even $\ge 2$ of all odd natural numbers,
there are for all other spherical spaces at low wave numbers
below a given threshold
$\beta_{\hbox{\scriptsize th}}$ ``gaps'' in the spectrum.
As can be seen from Table \ref{Tab:Spectrum}, the threshold value
$\beta_{\hbox{\scriptsize th}}$ increases if the volume $V({\cal M})$
decreases,
i.\,e.\ $\beta_{\hbox{\scriptsize th}}=4[(m+1)/2]+1$, 13, 25 and 61 for the
spaces belonging to the groups $D^\star_{4m}$, $T^\star$, $O^\star$ and
$I^\star$, respectively.
This demonstrates clearly the important r\^ole played by the spatial volume,
i.\,e.\ the topology of the Universe.

%\begin{landscape}
\begin{table}
\hspace*{-2cm}\begin{tabular}{|c|c|c|}
\hline
$\Gamma$ & wave number spectrum $\{\beta\}$ of manifold ${\cal M} = {\cal S}^3/\Gamma$ &
multiplicity $r^{\cal M}(\beta)$ \\
\hline
$Z_1$ & $\mathbb{N}$  & $\beta^2$ \\
\hline
$Z_m$, $m$ odd $\ge 1$ & $\{1,3,5,\dots,m\} \cup \{n|n\ge m+1\}$ &
$\beta\sum_{\beta-1\equiv 2l(m); 0\le l \le \beta-1} 1$ \\
\hline
$Z_m$, $m$ even $\ge 2$ & $2\mathbb{N}+1$ &
$\beta\sum_{\beta-1\equiv 2l(m); 0\le l \le \beta-1} 1$ \\
\hline
$D_{4m}^\star$, $m\ge 2$ &
$\{1,5,9,\dots,4\left[\frac{m+1}2\right]+1\} \cup
\{2n+1| n\ge 2\left[\frac{m+1}2\right]+1\}$ &
$\beta \left(\left[\frac{\beta-1}{2m}\right]+1\right)$ for
$\beta \in \{4n+1|n\ge 0\}$
\\ & &
$\beta \left[\frac{\beta-1}{2m}\right]$ for
$\beta \in \{4n+3|n\ge\left[\frac{m+1}2\right] \}$
\\
\hline
$T^\star$ & $\{1,7,9\} \cup \{2n+1|n\ge 6\}$ &
$\beta\left( 2 \left[\frac{\beta-1}6\right] + \left[\frac{\beta-1}4\right] -
\frac{\beta-3}2\right)$ \\
\hline
$O^\star$ & $\{1,9,13,17,19,21\} \cup \{2n+1| n\ge 12\}$ &
$\beta\left( \left[\frac{\beta-1}8\right] + \left[\frac{\beta-1}6\right] +
\left[\frac{\beta-1}4\right] - \frac{\beta-3}2\right)$ \\
\hline
$I^\star$ & $\{1,13,21,25,31,33,37,41,43,45,49,51,53,55,57\}$ &
$\beta\left( \left[\frac{\beta-1}{10}\right] + \left[\frac{\beta-1}6\right] +
\left[\frac{\beta-1}4\right] - \frac{\beta-3}2 \right)$ \\
& $\cup \, \{2n+1, n \ge 30\}$ & \\
\hline
\end{tabular}
\caption{\label{Tab:Spectrum}
The eigenvalue spectrum for the groups $\Gamma$.
}
\end{table}
%\end{landscape}

In addition to the wave number gaps, there is another important imprint
of topology on the multipole moments $C_l$ due to the multiplicities
$r^{\cal M}(\beta)$, also given in Table \ref{Tab:Spectrum},
by which the vibrational modes are weighted in the mode sum (\ref{Eq:Cl_NSW}).
Again, this effect is significant for the low vibrational modes and
thus for the low multipole moments which can be considered as carrying the
fingerprints of the topology of the Universe.

On the other hand, for the large multipole moments, 
$l \gg \beta_{\hbox{\scriptsize th}}$,
the details of the different topologies get washed out due to the
identical mean  asymptotic behaviour of the multiplicities for all
multi-connected spaces ${\cal M}$ (except $Z_m$, $m$ odd) given by
\begin{equation}
\label{Eq:multiplicity_asymptotic}
r^{\cal M}(\beta) \; = \;
\frac{2}{N} \, \beta^2 \, + \, \dots \; = \;
\frac{V({\cal M})}{\pi^2} \, \beta^2 \, + \, \dots
\hspace{10pt} \hbox{ for } \hspace{10pt}
\beta \to \infty
\end{equation}
which leads to the universal formula ($l \gg \beta_{\hbox{\scriptsize th}}$)
\begin{equation}
\label{Eq:Cl_NSW_asymptotic}
C_l^{\hbox{\scriptsize SW}}({\cal M}) \; \simeq \;
\frac{2}{9} \sum_{\beta {\,\hbox{\scriptsize odd}}, \beta>l}^\infty \;
P_\Phi(\beta) \, g_\beta^2(\eta_{\hbox{\scriptsize SLS}}) \,
R_{\beta l}^2(\tau_{\hbox{\scriptsize SLS}})
\hspace{10pt} .
\end{equation}
For the homogeneous lens spaces $Z_m$, $m$ odd $\ge 1$,
one obtains instead
\begin{equation}
\label{Eq:multiplicity_asymptotic_Zm_odd}
r^{\cal M}(\beta) \; = \;
\frac{1}{N} \, \beta^2 \, + \, \dots
\hspace{10pt} \hbox{ for } \hspace{10pt}
\beta \to \infty
\hspace{10pt} ,
\end{equation}
and thus from (\ref{Eq:Cl_NSW}) for $l \gg \beta_{\hbox{\scriptsize th}}$
\begin{equation}
\label{Eq:Cl_NSW_asymptotic_Zm_odd}
C_l^{\hbox{\scriptsize SW}}(Z_m) \; \simeq \;
\frac{1}{9} \sum_{\beta=l+1}^\infty \;
P_\Phi(\beta) \, g_\beta^2(\eta_{\hbox{\scriptsize SLS}}) \,
R_{\beta l}^2(\tau_{\hbox{\scriptsize SLS}})
\end{equation}
which should be numerically not very different from
the expression (\ref{Eq:Cl_NSW_asymptotic})
where the additional factor 2 should recompensate for the missing
even $\beta$-values in the sum (\ref{Eq:Cl_NSW_asymptotic}).
The asymptotic behaviour given in eqs.\,(\ref{Eq:multiplicity_asymptotic})
and (\ref{Eq:multiplicity_asymptotic_Zm_odd}) is in agreement with
Weyl's law, as can be seen as follows $(k\to\infty)$:
\begin{eqnarray} \nonumber
{\cal N}^{\cal M}(k) & := & \#\{\beta| \beta \leq k \} \; = \;
{\sum_{\beta\le k}} ' r^{\cal M}(\beta)
\\ & = &
\label{Eq:Weyl_gen}
\sum_{\beta {\,\hbox{\scriptsize odd}}, \beta\le k}
\left(\frac 2N \beta^2\right)
\, + \, \dots \; = \;
\frac{V({\cal M})}{6\pi^2} \, k^3 \, + \, \dots
\end{eqnarray}
for $\Gamma \neq Z_m, m$ odd $\ge 1$, and
\begin{equation}
\label{Eq:Weyl_Zm_odd}
{\cal N}^{\cal M}(k) \; = \;
\sum_{\beta=1}^k \left(\frac 1N \beta^2\right)
\, + \, \dots \; = \;
\frac{V({\cal M})}{6\pi^2} \, k^3 \, + \, \dots
\end{equation}
for $\Gamma = Z_m, m$ odd $\ge 1$.
One thus sees that the ordinary Sachs-Wolfe contribution
to the large multipoles is identical for all spherical spaces,
including ${\cal S}^3$, in agreement with the expectation
that the cosmic topology is most clearly seen at large scales.

The above discussion showed how the topology of the Universe influences
via the vibrational modes the CMB anisotropy at large scales.
The whole story is, however, more subtle, since an additional
$l$-dependence comes in eq.(\ref{Eq:Cl_NSW}) from the radial functions
$R_{\beta l}(\tau_{\hbox{\scriptsize SLS}})$, and furthermore,
there is an important dependence of the multipoles on the
curvature radius  respectively on $\Omega_{\hbox{\scriptsize tot}}$
which is determined by both the time evolution via
$g_\beta(\eta_{\hbox{\scriptsize SLS}})$ and the radial functions.
The interplay of all these effects is responsible for the rather
complicated structure displayed by the numerical computations
to be illustrated below.

\begin{figure}[htb]  %  TETRAEDER
\vspace*{-2.5cm}
\begin{center}
%\hspace*{-15pt}
\hspace*{-80pt}\begin{minipage}{14cm}
\begin{minipage}{6cm}
\includegraphics[width=9.0cm]{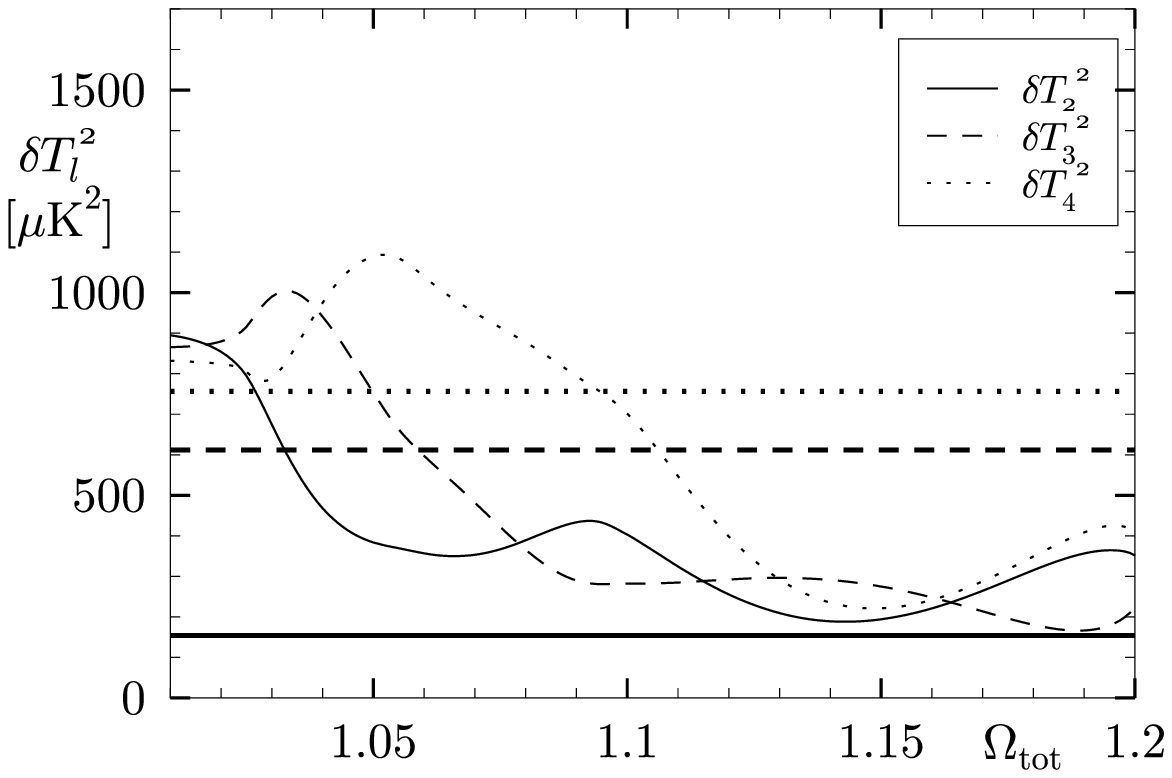}
\put(-180,155){(a)}
\end{minipage}
\begin{minipage}{6cm}
\hspace*{50pt}\includegraphics[width=9.0cm]{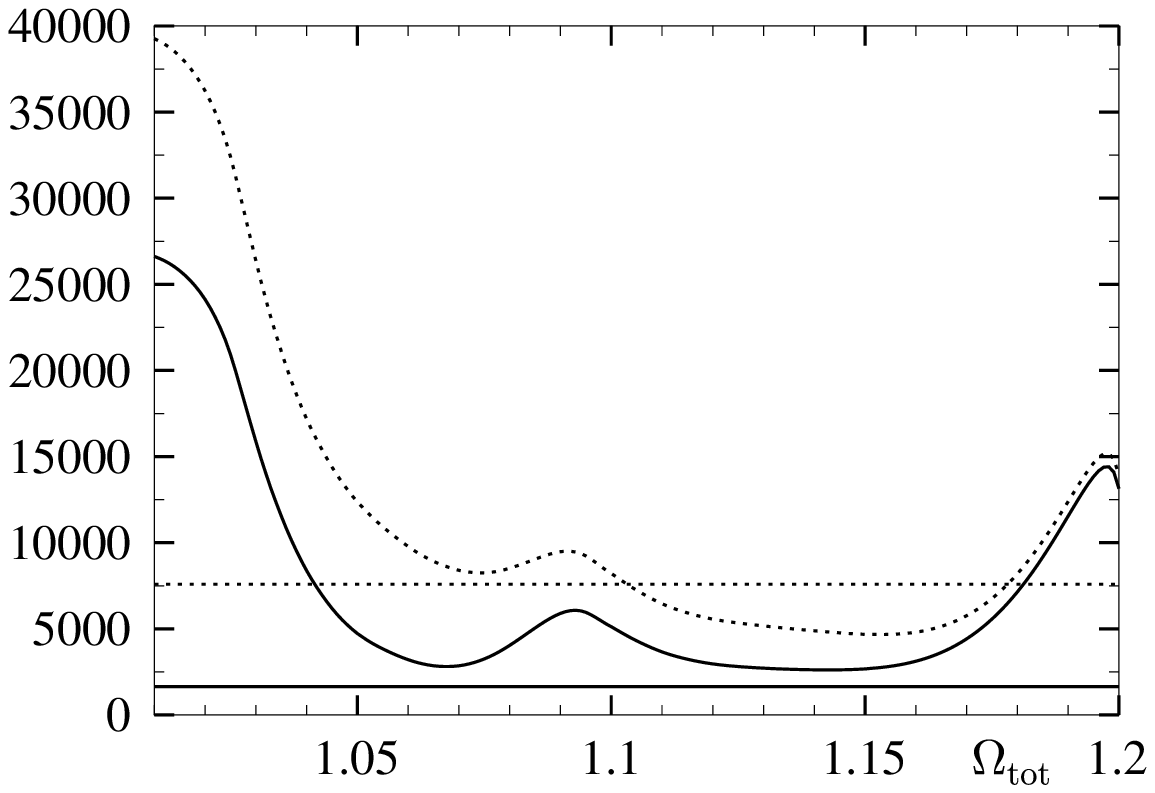}
\put(-175,155){(b)}
\end{minipage}
\end{minipage}
\end{center}
\vspace*{-1.3cm}
\caption{\label{Fig:Statistic_Tetraeder}
The binary tetrahedral group $T^\star$.
Panel (a) shows the $\Omega_{\hbox{\scriptsize tot}}$ dependence
of the mean value of the first three angular power moments $\delta T_l^2$.
The horizontal lines indicate the corresponding WMAP values for $\delta T_l^2$.
Panel (b) displays the $S(\rho)$ statistics
($\rho=60^\circ$ full curve and $\rho=20^\circ$ dotted curve) 
in units of $\mu\hbox{K}^4$
in dependence on $\Omega_{\hbox{\scriptsize tot}}$.
The corresponding WMAP values are indicated as horizontal lines.
}
\end{figure}

\begin{figure}[htb]  %  OKTAEDER
\vspace*{-2.5cm}
\begin{center}
%\hspace*{-15pt}
\hspace*{-80pt}\begin{minipage}{14cm}
\begin{minipage}{6cm}
\includegraphics[width=9.0cm]{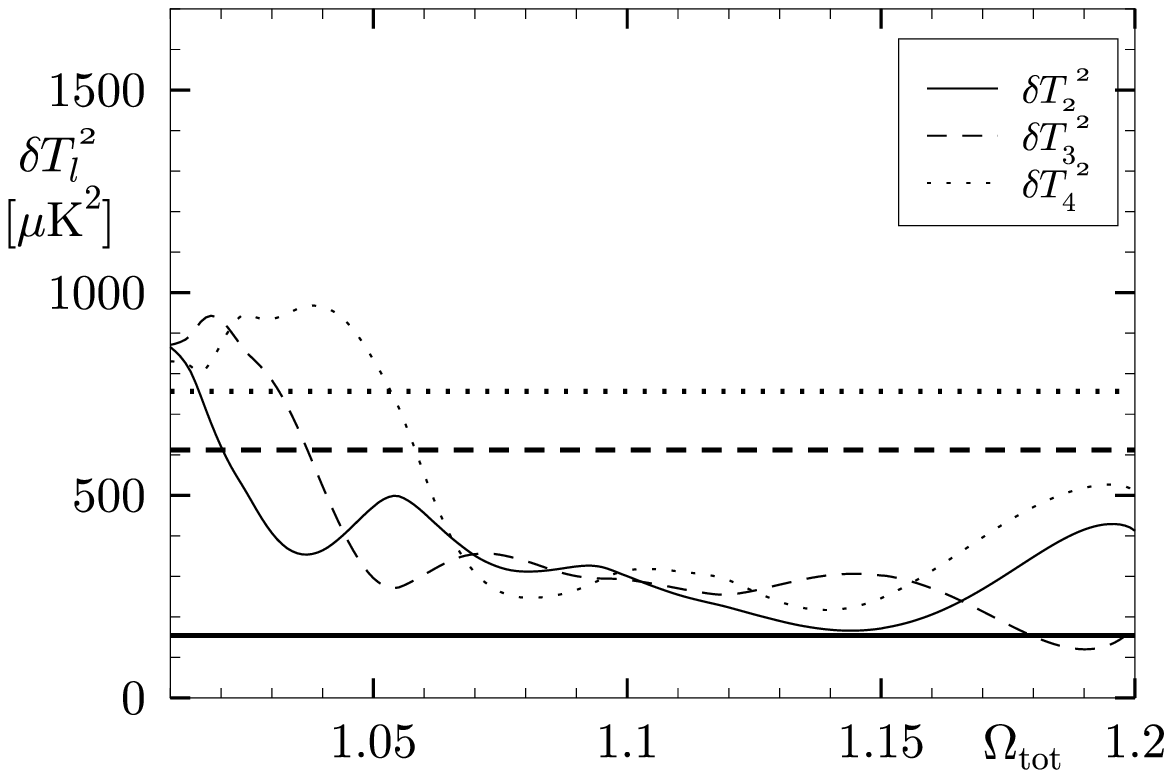}
\put(-180,155){(a)}
\end{minipage}
\begin{minipage}{6cm}
\hspace*{50pt}\includegraphics[width=9.0cm]{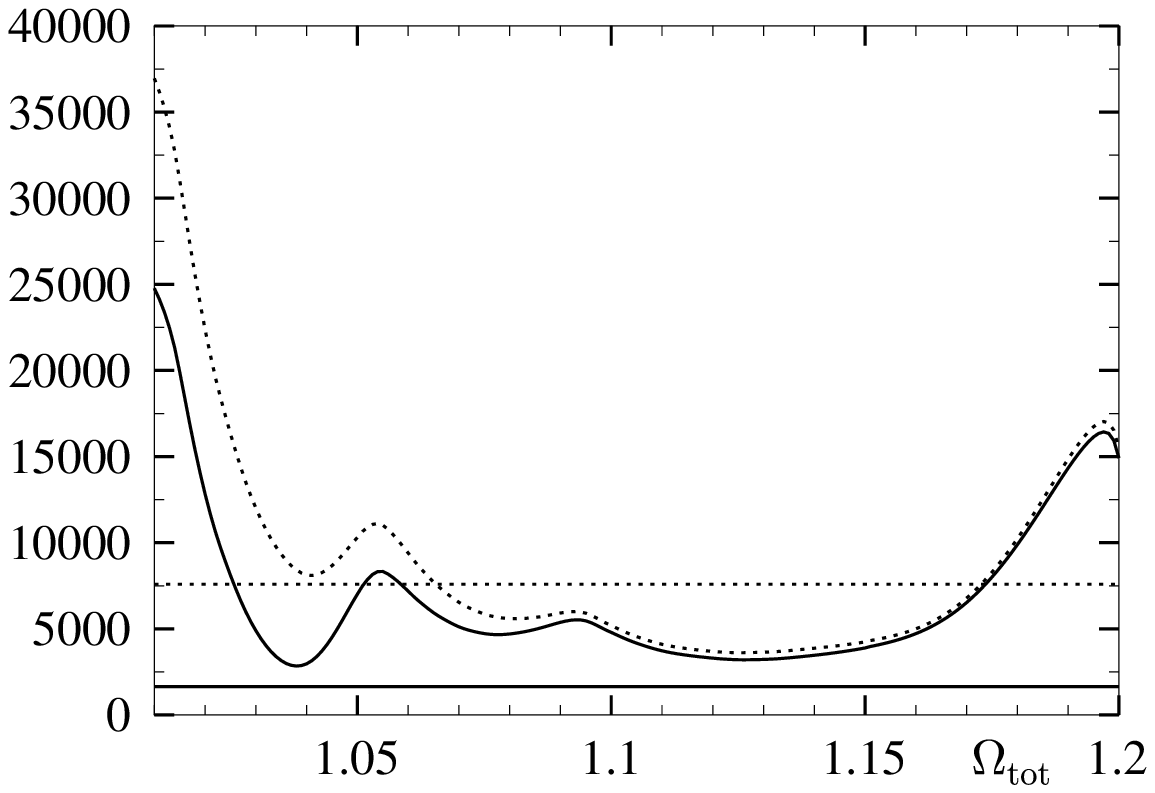}
\put(-175,155){(b)}
\end{minipage}
\end{minipage}
\end{center}
\vspace*{-1.3cm}
\caption{\label{Fig:Statistic_Oktaeder}
The same as in figure \ref{Fig:Statistic_Tetraeder}
for the binary octahedral group $O^\star$.
}
\end{figure}

\begin{figure}[htb]  %  DODEKAEDER
\vspace*{-2.5cm}
\begin{center}
%\hspace*{-15pt}
\hspace*{-80pt}\begin{minipage}{14cm}
\begin{minipage}{6cm}
\includegraphics[width=9.0cm]{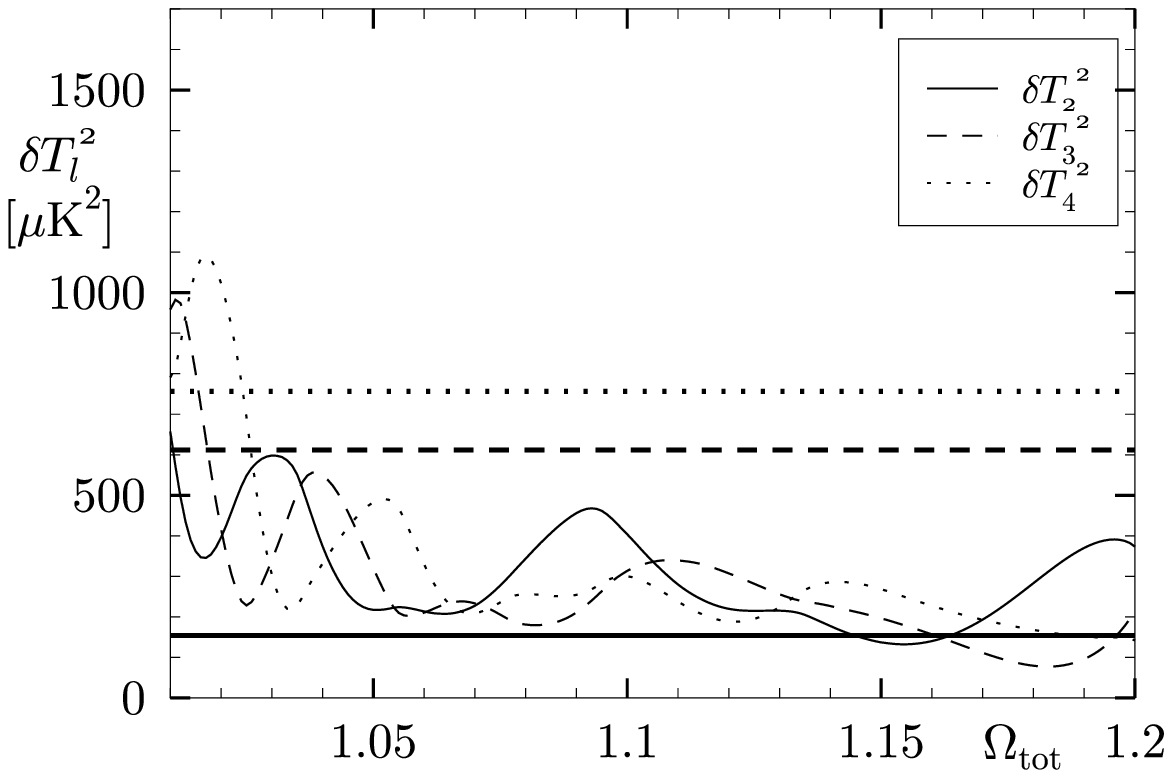}
\put(-180,155){(a)}
\end{minipage}
\begin{minipage}{6cm}
\hspace*{50pt}\includegraphics[width=9.0cm]{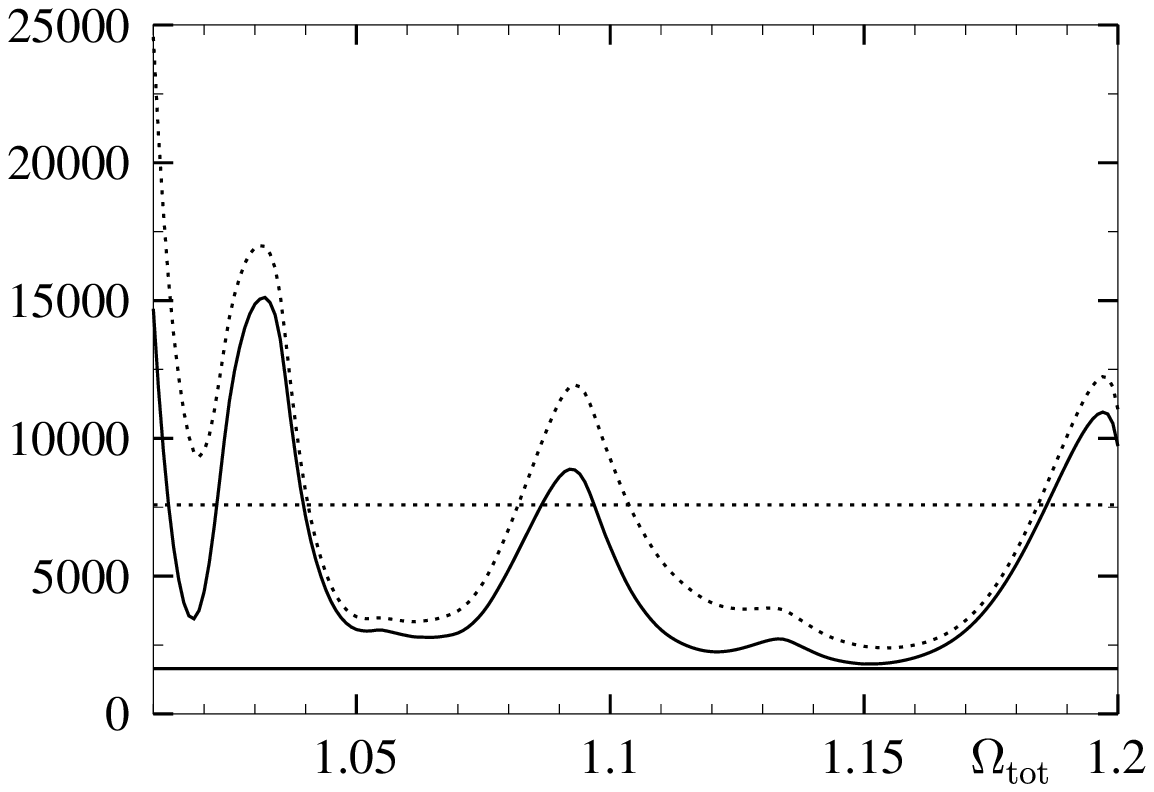}
\put(-175,155){(b)}
\end{minipage}
\end{minipage}
\end{center}
\vspace*{-1.3cm}
\caption{\label{Fig:Statistic_Dodekaeder}
The same as in figure \ref{Fig:Statistic_Tetraeder}
for the binary icosahedral group $I^\star$.
}
\end{figure}

\begin{figure}[htb]  %  DIHEDRAL 4X2
\vspace*{-2.5cm}
\begin{center}
%\hspace*{-15pt}
\hspace*{-80pt}\begin{minipage}{14cm}
\begin{minipage}{6cm}
\includegraphics[width=9.0cm]{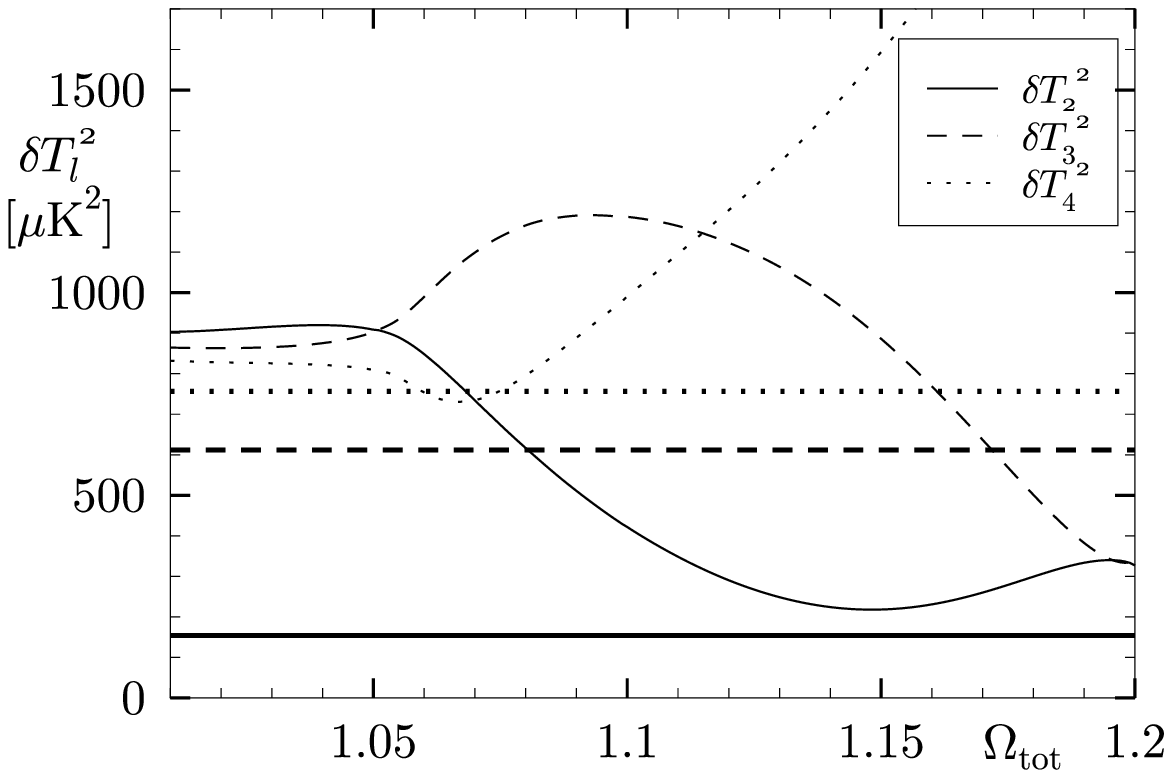}
\put(-180,155){(a)}
\end{minipage}
\begin{minipage}{6cm}
\hspace*{50pt}\includegraphics[width=9.0cm]{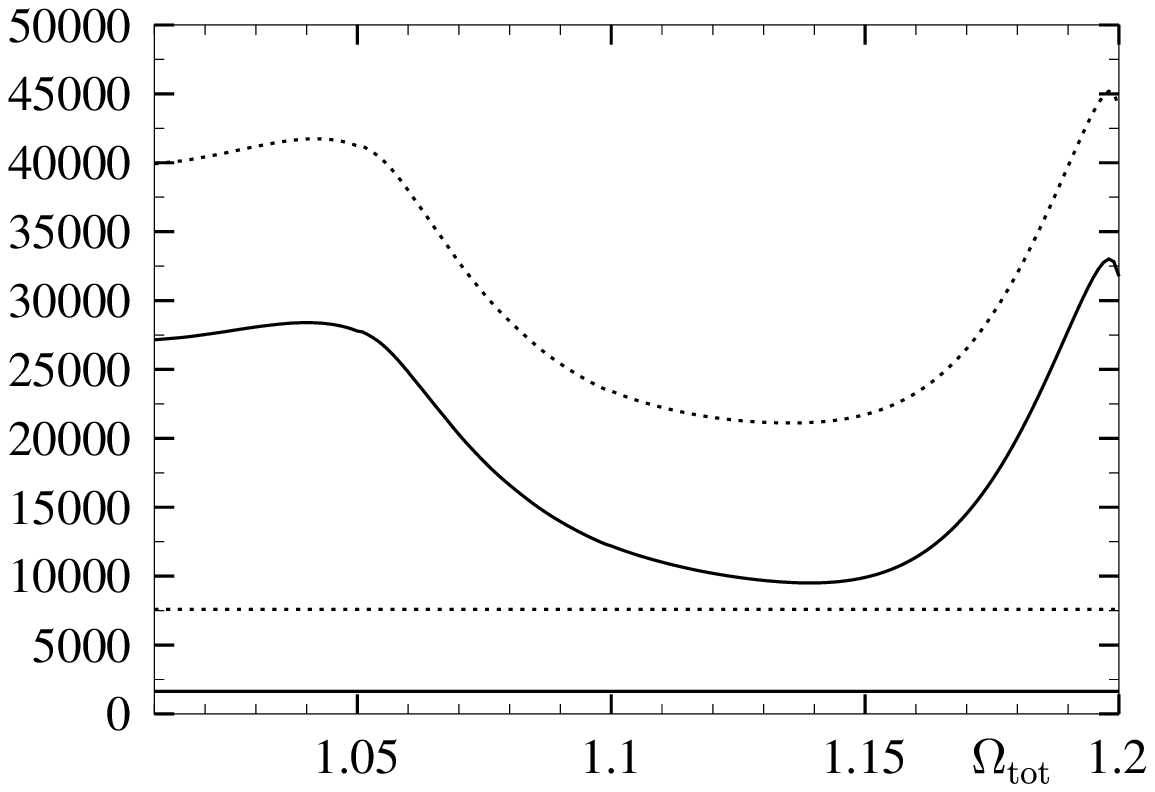}
\put(-175,155){(b)}
\end{minipage}
\end{minipage}
\end{center}
\vspace*{-1.3cm}
\caption{\label{Fig:Statistic_dihedral_4x2}
The same as in figure \ref{Fig:Statistic_Tetraeder}
for the binary dihedral group $D^\star_8$, i.\,e.\ $m=2$.
}
\end{figure}

\begin{figure}[htb]  %  DIHEDRAL 4X5
\vspace*{-2.5cm}
\begin{center}
%\hspace*{-15pt}
\hspace*{-80pt}\begin{minipage}{14cm}
\begin{minipage}{6cm}
\includegraphics[width=9.0cm]{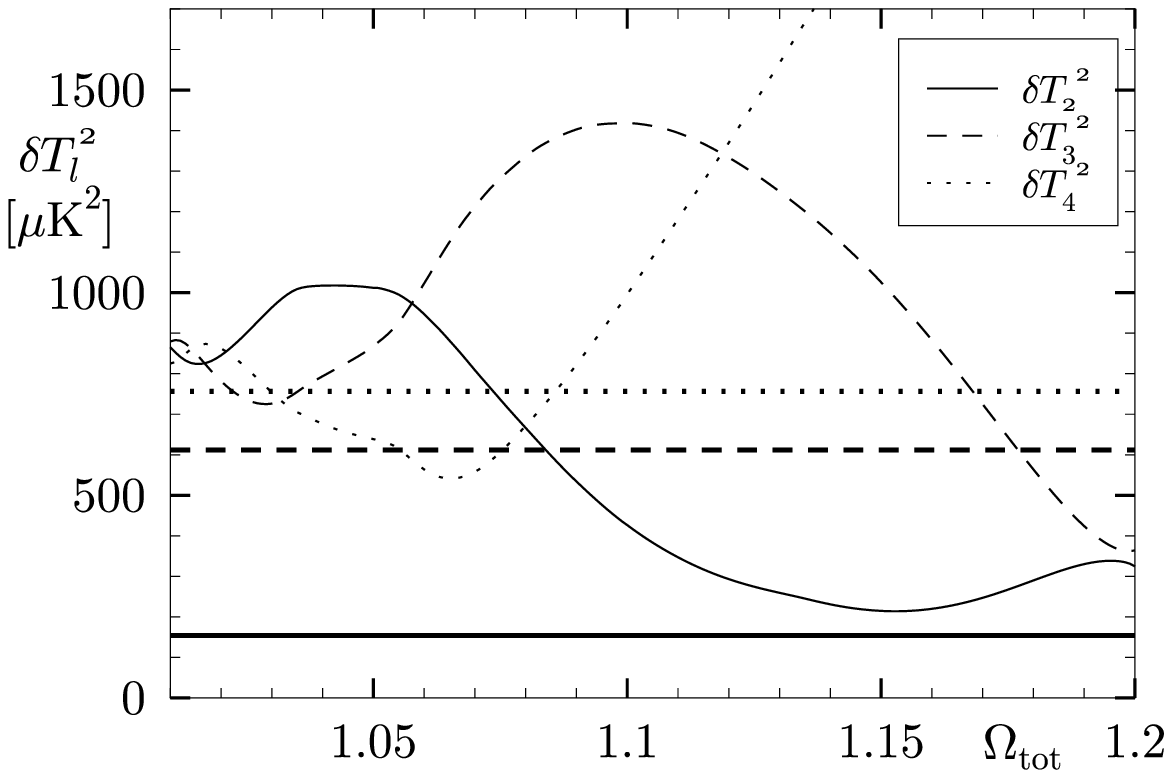}
\put(-180,155){(a)}
\end{minipage}
\begin{minipage}{6cm}
\hspace*{50pt}\includegraphics[width=9.0cm]{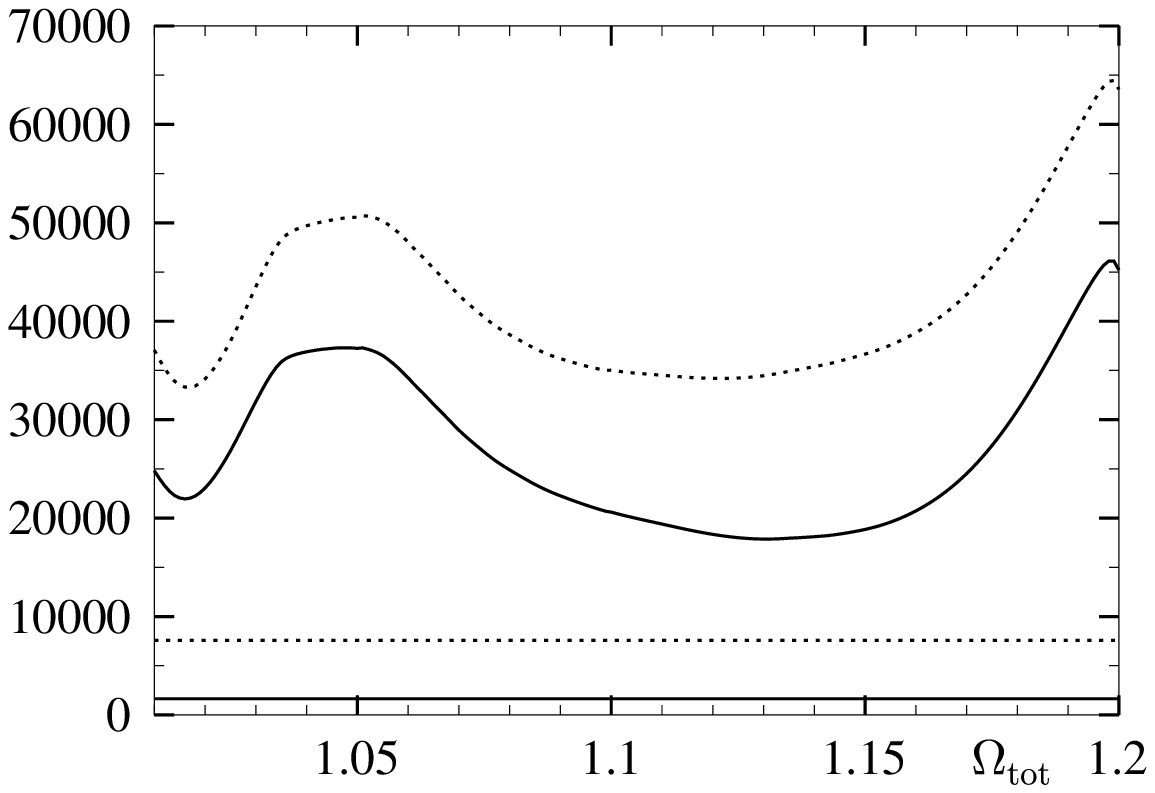}
\put(-175,155){(b)}
\end{minipage}
\end{minipage}
\end{center}
\vspace*{-1.3cm}
\caption{\label{Fig:Statistic_dihedral_4x5}
The same as in figure \ref{Fig:Statistic_Tetraeder}
for the binary dihedral group $D^\star_{20}$, i.\,e.\ $m=5$.
}
\end{figure}

\begin{figure}[htb]  %  ZYKLISCH Z_1  (S3)
\vspace*{-2.5cm}
\begin{center}
%\hspace*{-15pt}
\hspace*{-80pt}\begin{minipage}{14cm}
\begin{minipage}{6cm}
\includegraphics[width=9.0cm]{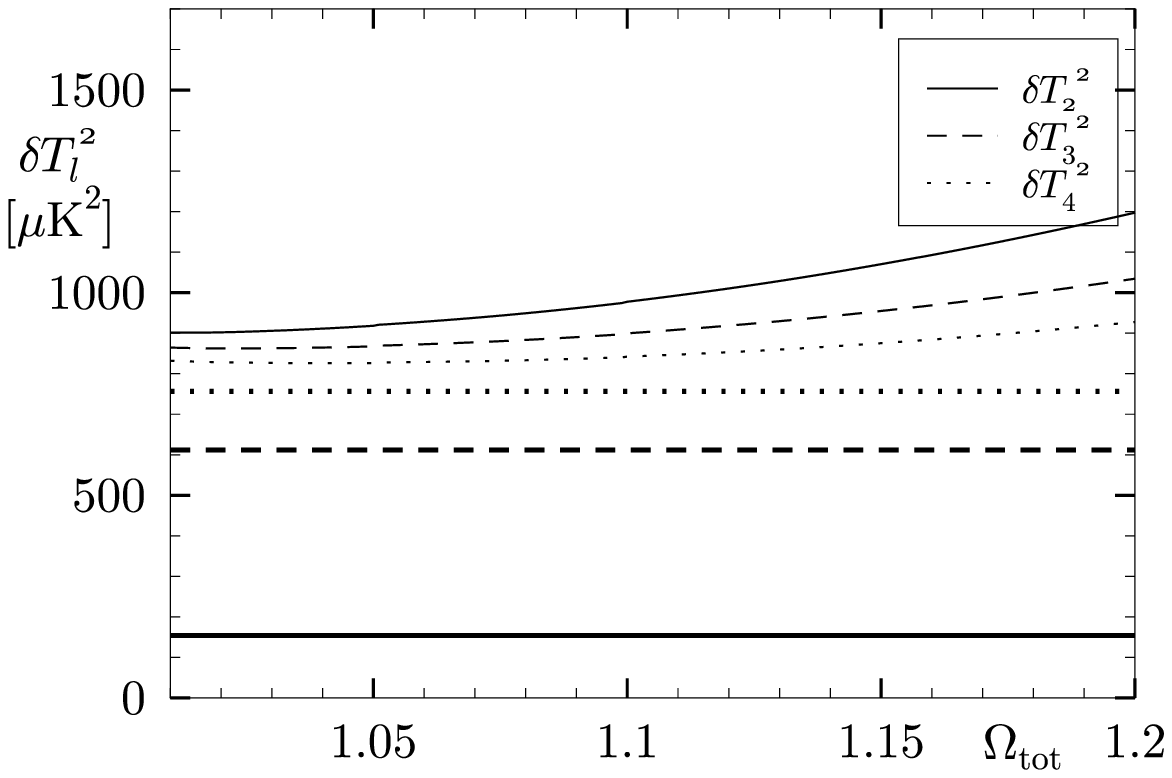}
\put(-180,155){(a)}
\end{minipage}
\begin{minipage}{6cm}
\hspace*{50pt}\includegraphics[width=9.0cm]{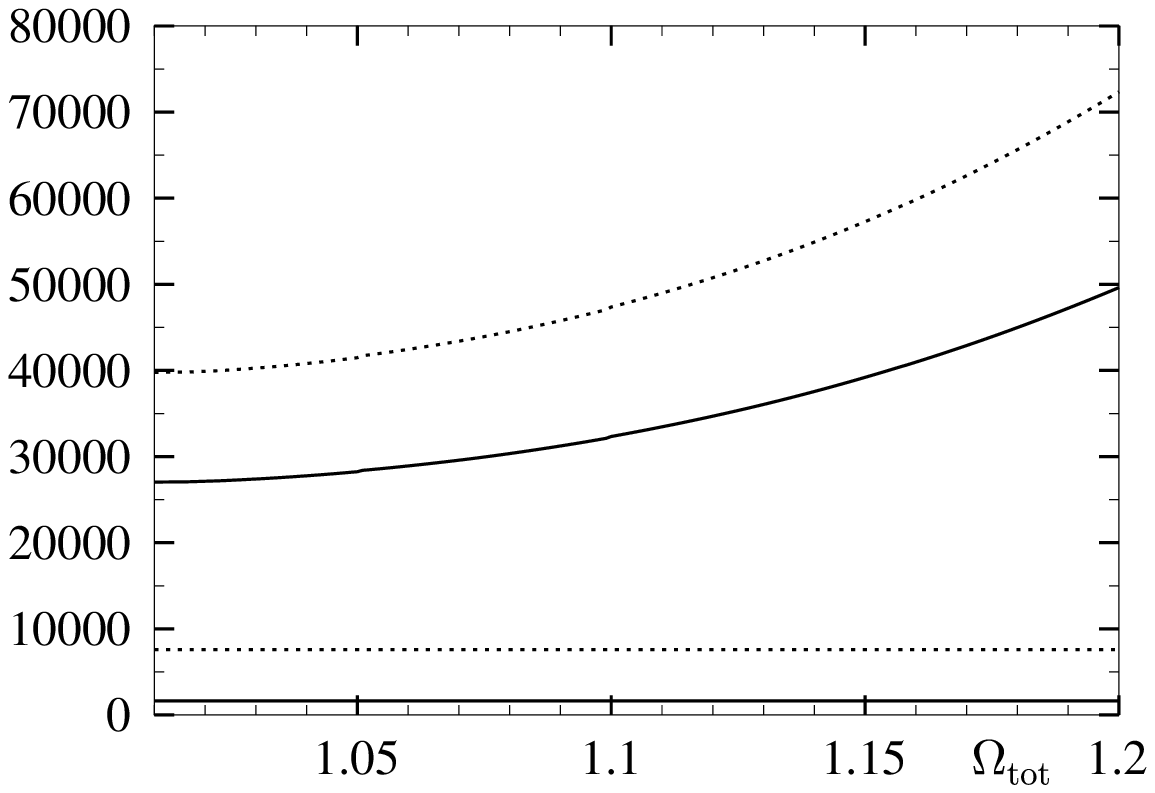}
\put(-175,155){(b)}
\end{minipage}
\end{minipage}
\end{center}
\vspace*{-1.3cm}
\caption{\label{Fig:Statistic_zyklisch_1}
The same as in figure \ref{Fig:Statistic_Tetraeder}
for the cyclic group $Z_1$, i.\,e.\ for the space ${\cal S}^3$.
}
\end{figure}

\begin{figure}[htb]  %  ZYKLISCH Z_2  (P3)
\vspace*{-2.5cm}
\begin{center}
%\hspace*{-15pt}
\hspace*{-80pt}\begin{minipage}{14cm}
\begin{minipage}{6cm}
\includegraphics[width=9.0cm]{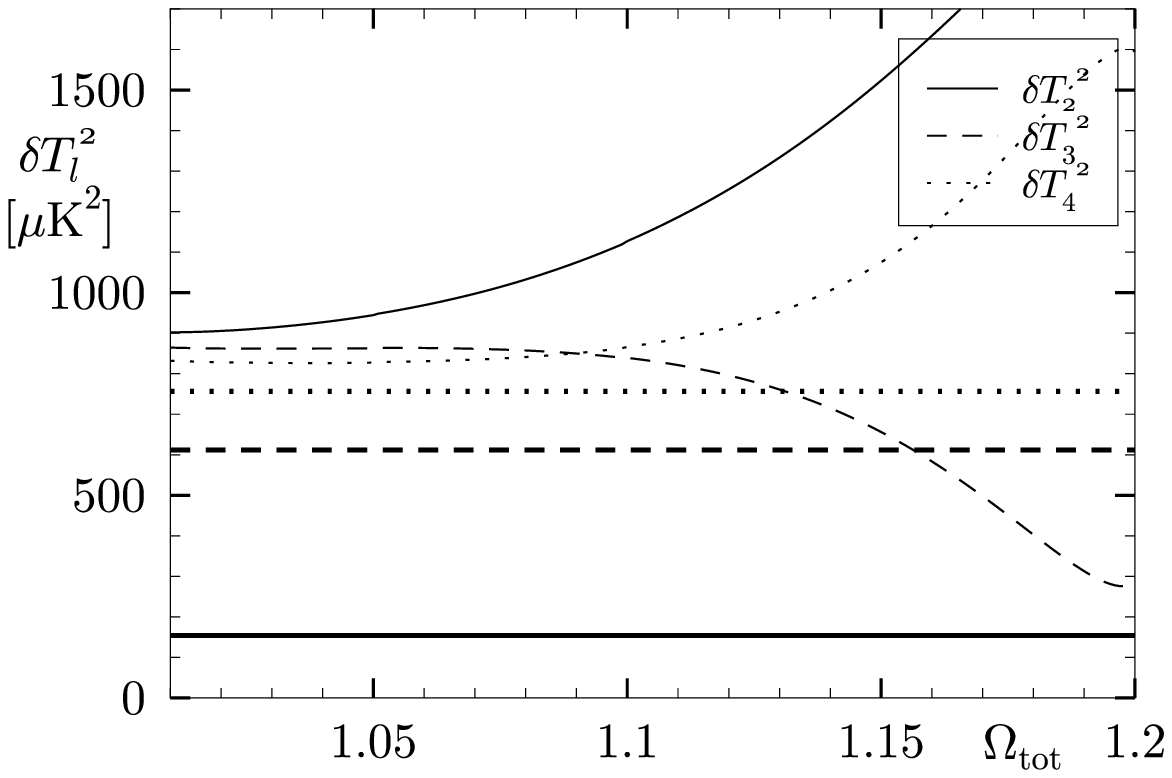}
\put(-180,155){(a)}
\end{minipage}
\begin{minipage}{6cm}
\hspace*{50pt}\includegraphics[width=9.0cm]{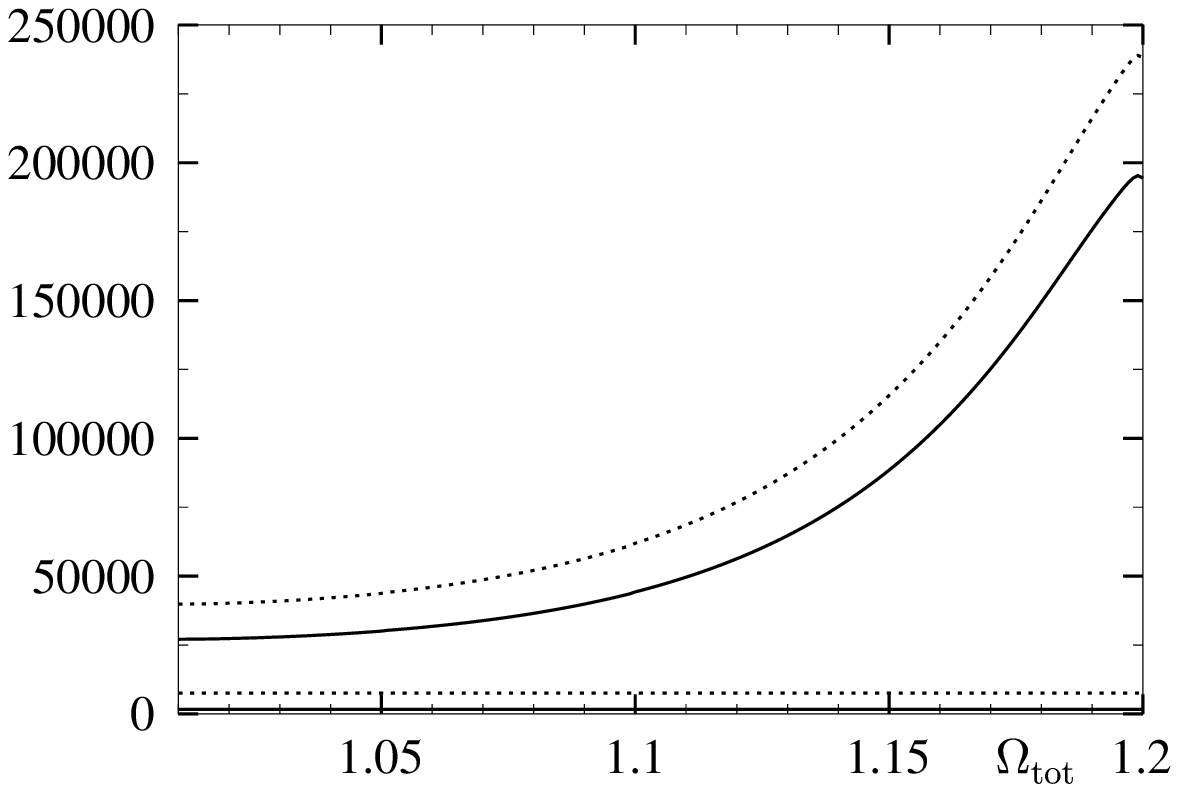}
\put(-175,155){(b)}
\end{minipage}
\end{minipage}
\end{center}
\vspace*{-1.3cm}
\caption{\label{Fig:Statistic_zyklisch_2}
The same as in figure \ref{Fig:Statistic_Tetraeder}
for the cyclic group $Z_2$, i.\,e.\ for the projective space ${\cal P}^3$.
}
\end{figure}

\begin{figure}[htb]  %  ZYKLISCH Z_6
\vspace*{-2.5cm}
\begin{center}
%\hspace*{-15pt}
\hspace*{-80pt}\begin{minipage}{14cm}
\begin{minipage}{6cm}
\includegraphics[width=9.0cm]{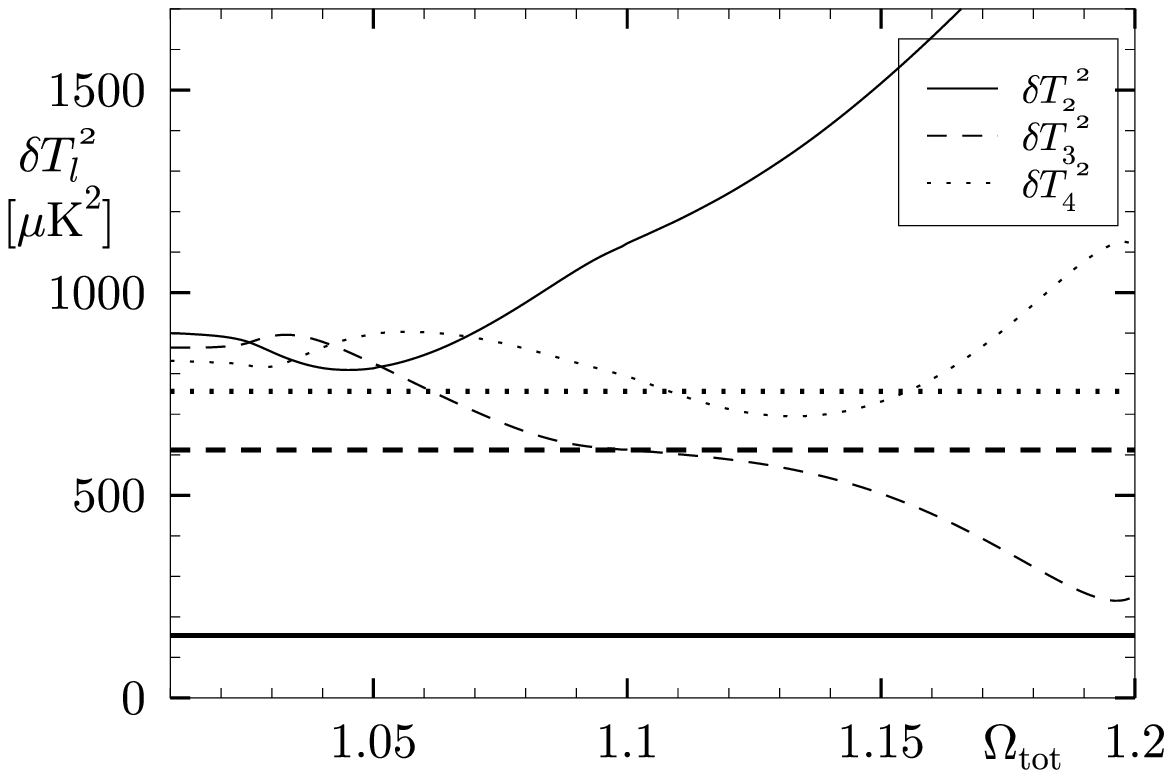}
\put(-180,155){(a)}
\end{minipage}
\begin{minipage}{6cm}
\hspace*{50pt}\includegraphics[width=9.0cm]{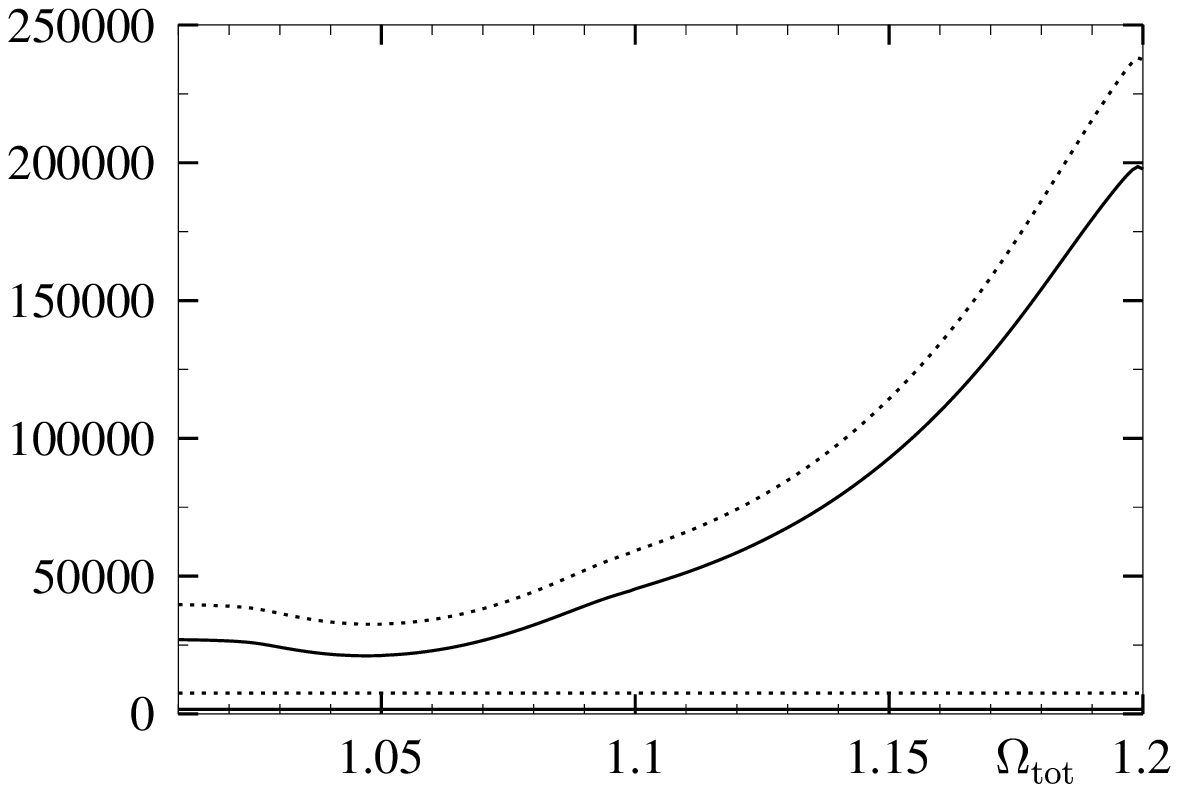}
\put(-175,155){(b)}
\end{minipage}
\end{minipage}
\end{center}
\vspace*{-1.3cm}
\caption{\label{Fig:Statistic_zyklisch_6}
The same as in figure \ref{Fig:Statistic_Tetraeder}
for the cyclic group $Z_6$.
}
\end{figure}

Let us now discuss the results for the different spherical space forms.
In all these computations the CMB anisotropy is obtained using the complete
Sachs-Wolfe formula $(\tau(\eta):=\eta_0 - \eta)$
\begin{eqnarray} \nonumber \hspace{-50pt}
\frac{\delta T}T(\hat n) & \hspace*{-15pt}= &
{\sum_{\beta\ge 3}} '\; \sum_{i=1}^{r^{\cal M}(\beta)} \; 
\left[ \left( \Phi_\beta^i(\eta) +
\frac{\delta_{\gamma,\beta}^i(\eta)}4 +
\frac{a(\eta) V_{\gamma,\beta}^i(\eta)}{E_\beta}
\frac{\partial}{\partial \tau} \right)
\Psi_\beta^{{\cal M},i}(\tau(\eta),\theta,\phi)
\right]_{\eta=\eta_{\hbox{\scriptsize{SLS}}}}
\\ & & \hspace{10pt}
\label{Eq:Sachs_Wolfe_tight_coupling}
\, + \,
2 \; {\sum_{\beta\ge 3}} '\; \sum_{i=1}^{r^{\cal M}(\beta)} \; 
\int_{\eta_{\hbox{\scriptsize{SLS}}}}^{\eta_0} d\eta \,
\frac{\partial\Phi_\beta^i(\eta)}{\partial\eta} \,
\Psi_\beta^{{\cal M},i}(\tau(\eta),\theta,\phi)
\hspace{10pt} .
\end{eqnarray}
The details of the computation of the quantities needed in
(\ref{Eq:Sachs_Wolfe_tight_coupling}) are described in
\cite{Aurich_Lustig_Steiner_Then_2004a}.
The first two terms in (\ref{Eq:Sachs_Wolfe_tight_coupling})
are the ordinary Sachs-Wolfe contribution (\ref{Eq:NSW})
discussed above.
The next term involving the spatial covariant divergence of
the velocity field is the Doppler contribution.
The integral over the photon path yields the
integrated Sachs-Wolfe contribution. 
The $\beta_{\hbox{\scriptsize max}}$ cut-off is chosen sufficiently
high in order to obtain enough
multipoles which then enables us to normalise the $\delta T_l^2$ spectrum
in the range $l=20$ to 45 according to the WMAP first-year data
\cite{Bennett_et_al_2003}.
In our computations we use
$\beta_{\hbox{\scriptsize max}}=3001$ for
$\Omega_{\hbox{\scriptsize tot}} \leq 1.05$,
$\beta_{\hbox{\scriptsize max}}=2001$ for
$1.05 < \Omega_{\hbox{\scriptsize tot}} \leq 1.1$, and
$\beta_{\hbox{\scriptsize max}}=1501$ for
$1.1 < \Omega_{\hbox{\scriptsize tot}} \leq 1.2$.
In the following we fix the cosmological parameters as
$\Omega_{\hbox{\scriptsize bar}} = 0.046$,
$\Omega_{\hbox{\scriptsize mat}} = 0.28$, and $h=70$,
in agreement with the WMAP data.
The free parameter is
$\Omega_\Lambda = \Omega_{\hbox{\scriptsize tot}} -
\Omega_{\hbox{\scriptsize mat}}-\Omega_{\hbox{\scriptsize rad}}$,
i.\,e.\ the density corresponding to the cosmological constant.
Although $\Omega_\Lambda$ is varied, we show in the following
figures $\Omega_{\hbox{\scriptsize tot}}$ on the abscissa.

In figures \ref{Fig:Statistic_Tetraeder} to \ref{Fig:Statistic_zyklisch_6}
we show the dependence of the large scale CMB anisotropy
on $\Omega_{\hbox{\scriptsize tot}}$ for
the binary tetrahedral space, the binary octahedral space,
the dodecahedral space, 
two binary dihedral spaces and three spaces belonging to cyclic groups.
In panel (a) the expectation values of the
angular power spectra $\delta T_l^2$ are shown
for the first three multipole moments $\delta T_2^2$ (solid curve),
$\delta T_3^2$ (dashed curve) and $\delta T_4^2$ (dotted curve).
(Panel (b) in figures \ref{Fig:Statistic_Tetraeder} to
\ref{Fig:Statistic_zyklisch_6} will be discussed below.)
The corresponding values measured by the WMAP team are indicated
as straight horizontal lines.
With respect to the large scale anisotropy,
the best values for $\Omega_{\hbox{\scriptsize tot}}$ are obtained
by choosing those values which yield the best agreement with the data.
For example, consider the binary tetrahedral space
shown in figure \ref{Fig:Statistic_Tetraeder}.
Since the density $\Omega_{\hbox{\scriptsize tot}}$ should be
as close to one as possible, one chooses the first minimum of $\delta T_2^2$,
i.\,e.\ the range $\Omega_{\hbox{\scriptsize tot}}=1.06\dots 1.07$.
In this range the value of $\delta T_3^2$ is also near to the observed one.
One should note that these values are only the expectation values such
that for a given realization one has also to take into account
the cosmic variance.
However, as the simulations for such realizations show,
the probability for a given spherical space increases significantly
when the expectation values are already near to the observed values.
In this way one gets for the binary octahedral space
(figure \ref{Fig:Statistic_Oktaeder})
the range $\Omega_{\hbox{\scriptsize tot}}=1.03\dots 1.04$
and for the dodecahedron (figure \ref{Fig:Statistic_Dodekaeder})
$\Omega_{\hbox{\scriptsize tot}}=1.015\dots 1.02$.

For none of the considered binary dihedral spaces is a comparably good
agreement found.
We have computed the CMB anisotropy for the binary dihedral groups
$D_{4m}^\star$ with $m=2$, 3, 4, 5, 10, 20, and 30.
In figures \ref{Fig:Statistic_dihedral_4x2} and
\ref{Fig:Statistic_dihedral_4x5} we show as two typical examples
the cases $m=2$ and $m=5$, respectively.
The first multipole moment $\delta T_2^2$ decreases for all
considered groups $D_{4m}^\star$ only for unrealistically high values of
$\Omega_{\hbox{\scriptsize tot}}>1.1$.
In this parameter range the values for $\delta T_4^2$ are very large.
Thus a binary dihedral topology seems to be very unprobable as a possibility
for our Universe.

The cyclic groups $Z_m$ are also much worse
compared to the binary tetrahedral space, the binary octahedral space,
or the dodecahedron.
We have computed all groups $Z_m$ with $m\le 20$ and
a lot of examples up to $m=500$.
As three examples we present in figures \ref{Fig:Statistic_zyklisch_1} to
\ref{Fig:Statistic_zyklisch_6} the models for $m=1$, 2, and 6, respectively.
The group $Z_1$ has only the identity as a group element and
thus leads to the usual spherical space ${\cal S}^3$.
One observes the known difficulty of the concordance model,
i.\,e.\ too large values for the two lowest multipole moments.
The next group shown is $Z_2$ leading to
the projective space ${\cal P}^3$, also known as elliptic space,
which has historically played a special role as an example of
an alternative to the spherical space ${\cal S}^3$.
For $\Omega_{\hbox{\scriptsize tot}}<1.1$ the first multipole moments
behave very similar to the former case with $m=1$
such that this topology is not a better match than the concordance model.
Models with larger groups $Z_m$ lead to a quadrupole moment
which increases with increasing $\Omega_{\hbox{\scriptsize tot}}$, in general.
As an example, figure \ref{Fig:Statistic_zyklisch_6} displays the
result for $Z_6$.
The quadrupole moment $\delta T_2^2$ shows a very small minimum around
$\Omega_{\hbox{\scriptsize tot}}=1.05$ and increases for higher
values of $\Omega_{\hbox{\scriptsize tot}}$.
Although there are minima in many cyclic models,
they never suppress the power of $\delta T_2^2$ even to the level of
the observed value for $\delta T_4^2$,
see figure \ref{Fig:Statistic_zyklisch_6}.
The fact that lens spaces do not fit well the CMB anisotropy
is already discussed in \cite{Uzan_Riazuelo_Lehoucq_Weeks_2003}
where this behaviour is ascribed to the non well-proportioned
fundamental cells.

Therefore, from all spherical spaces,
only the binary tetrahedral space, the binary octahedral space,
and the dodecahedron display for the given parameter ranges
the observed suppression in the large anisotropy power.

Up to now, we have only discussed the angular power spectrum $\delta T_l^2$.
Let us now come to the {\it temperature two-point correlation function}
$C(\vartheta)$, which is defined as
$C(\vartheta) := \left< \delta T(\hat n) \delta T(\hat n')\right>$
with $\hat n \cdot \hat n' = \cos\vartheta$.
It can be computed from the multipole moments (\ref{Eq:C_l})
under the assumption of statistical isotropy as
\begin{equation}
\label{Eq:C_theta}
C(\vartheta) \; \simeq \;
\frac 1{4\pi} \, \sum_{l=2}^\infty \, (2l+1) \, C_l \, P_l(\cos\vartheta)
\hspace{10pt} .
\end{equation}
This quantity is well suited in order to measure the large scale power
as emphasised in \cite{Spergel_et_al_2003}
where the observations are compared with the theoretical models.
A comparison of $C(\vartheta)$ observed by WMAP with the concordance model
can also be found in figure 1 of reference \cite{Aurich_Lustig_Steiner_2004c}.
(In \cite{Aurich_Lustig_Steiner_2004c} we have derived an analytic
expression for $C^{\hbox{\scriptsize SW}}(\vartheta)$,
i.\,e.\ for the SW contribution in the case of the dodecahedron,
which after multiplication by $N/120$ holds for all homogeneous
spherical space forms.)
The correlation function $C(\vartheta)$ displays a surprisingly low
CMB anisotropy on large angular scales $\vartheta \ge \rho$,
which can be quantified by the $S(\rho)$ statistic \cite{Bennett_et_al_2003}
\begin{equation}
\label{Eq:S-Statistik}
S(\rho) \; = \;
\int_{-1}^{\cos\rho} \big| C(\vartheta) \big|^2 \; d\cos\vartheta
\end{equation}
which is discussed for the first-year WMAP data in \cite{Spergel_et_al_2003}
for $\rho=60^\circ$,
and it is found that only $0.3\%$ of the simulations based on the
concordance model ri have lower values of $S(60^\circ)$
than the observed value $S(60^\circ)=1644$.
The $S(\rho)$ statistic is shown for the above discussed spherical spaces
in figures \ref{Fig:Statistic_Tetraeder}
to \ref{Fig:Statistic_zyklisch_6} in panel (b)
for $\rho=60^\circ$ (solid curves) and $\rho=20^\circ$ (dotted curves).
The corresponding WMAP values are indicated as straight horizontal lines.
The inspection of these figures leads to the same result as
the above discussion of the first three angular power moments $\delta T_l^2$.
The binary tetrahedral space possesses a sufficiently low power in the
range $\Omega_{\hbox{\scriptsize tot}}=1.06\dots 1.07$,
the binary octahedral space in the range
$\Omega_{\hbox{\scriptsize tot}}=1.03\dots 1.04$,
and the dodecahedral space in the range
$\Omega_{\hbox{\scriptsize tot}}=1.015\dots 1.02$.
For the other spherical spaces no comparable agreement is found
as can be seen in the case of the two binary dihedral spaces
(figures \ref{Fig:Statistic_dihedral_4x2}(b) and
\ref{Fig:Statistic_dihedral_4x5}(b))
and the three cyclic groups
(figures \ref{Fig:Statistic_zyklisch_1}(b) to
\ref{Fig:Statistic_zyklisch_6}(b)).

\begin{figure}[htb]  %  Radial-Funktion
\vspace*{-2.5cm}
\begin{center}
\hspace*{-45pt}
\hspace*{-80pt}\begin{minipage}{14cm}
\begin{minipage}{6cm}
\includegraphics[width=9.0cm]{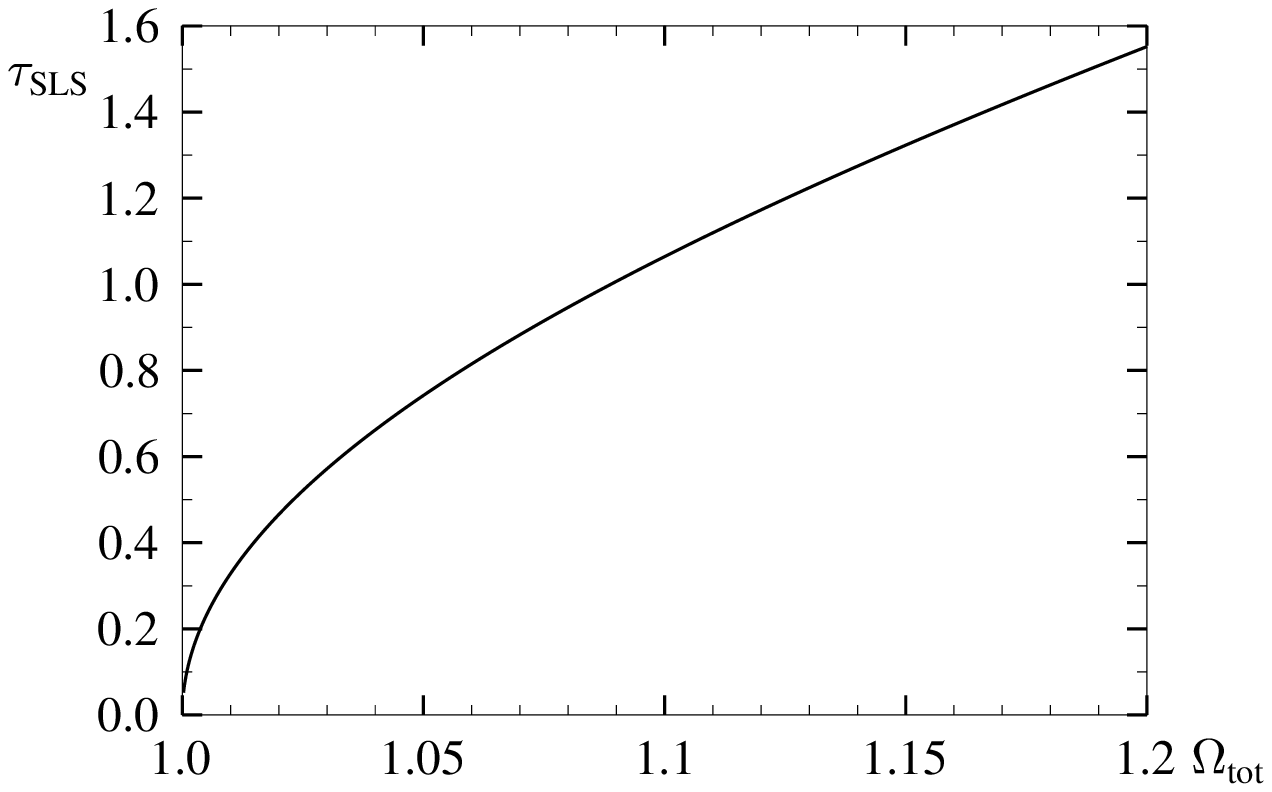}
\put(-180,155){(a)}
\end{minipage}
\begin{minipage}{6cm}
\hspace*{50pt}\includegraphics[width=9.0cm]{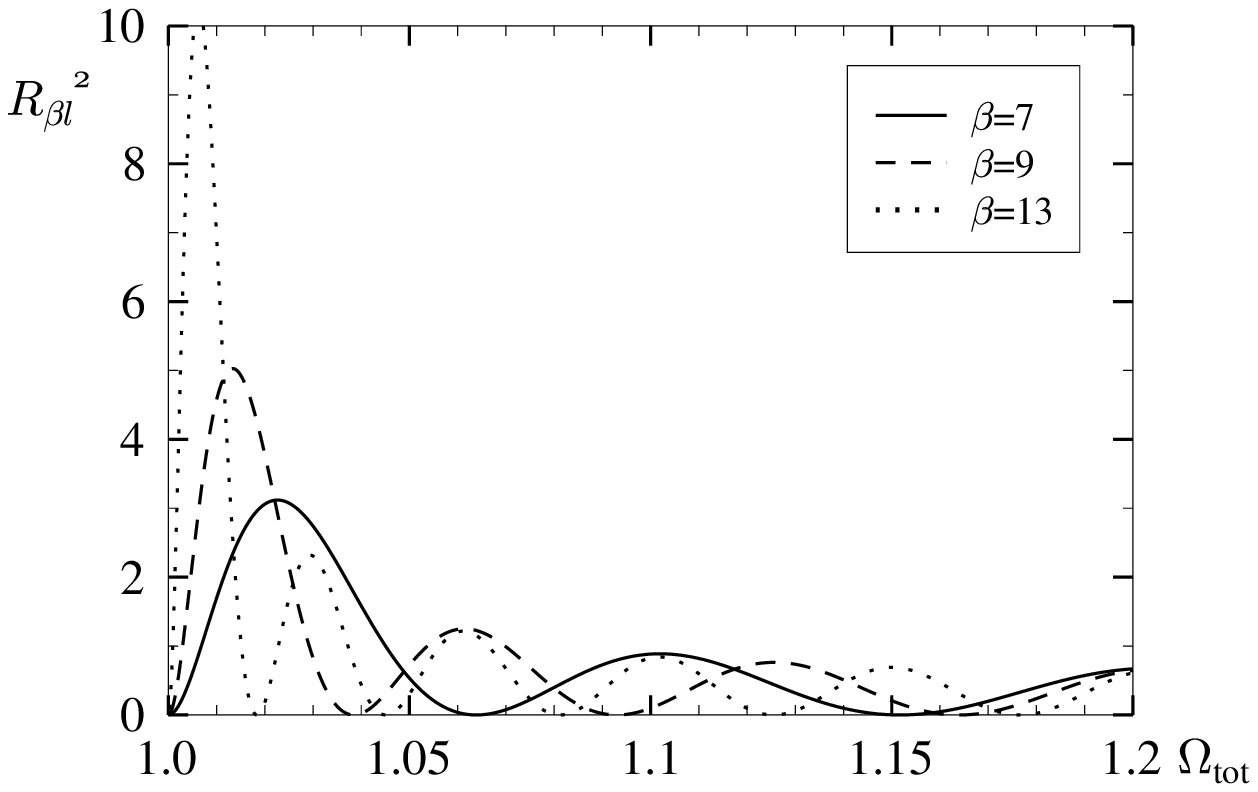}
\put(-150,155){(b)}
\end{minipage}
\end{minipage}
\end{center}
\vspace*{-1.3cm}
\caption{\label{Fig:Radialfunction}
Panel (a) shows the dependence of the conformal distance
$\tau_{\hbox{\scriptsize SLS}}$ to the surface of last scattering
on the density $\Omega_{\hbox{\scriptsize tot}}$.
Panel (b) shows the square of the radial function $R_{\beta l}^2(\tau)$
for the first eigenvalue occurring in (\ref{Eq:Sachs_Wolfe_tight_coupling})
for the quadrupole moment $l=2$
for the binary tetrahedral space $(\beta=7)$,
the binary octahedral space $(\beta=9)$,
and the dodecahedron $(\beta=13)$.
}
\end{figure}

Now, we would like to demonstrate the strong influence
of the radial function $R_{\beta l}^2(\tau)$
on the suppression of power in the case of the quadrupole moment $l=2$
already discussed above.
The quadrupole suppression gets stronger,
the higher the value of the first contributing eigenvalue is,
see equation (\ref{Eq:Cl_NSW}).
In addition, it is seen from equation (\ref{Eq:Cl_NSW})
that the term belonging to the first eigenvalue $\beta$ is multiplied
by $R_{\beta l}^2(\tau)$.
Thus the contribution of the first eigenvalue is eliminated
for those values of $\Omega_{\hbox{\scriptsize tot}}$
which belong to a value of $\tau_{\hbox{\scriptsize SLS}}$
at which $R_{\beta l}(\tau_{\hbox{\scriptsize SLS}})$ is zero.
The dependence of $\tau_{\hbox{\scriptsize SLS}}$ on
the density $\Omega_{\hbox{\scriptsize tot}}$ is shown
in figure \ref{Fig:Radialfunction}(a)
for our choice of cosmological parameters
and is well described by
$\tau_{\hbox{\scriptsize SLS}} =
3.32 \sqrt{\Omega_{\hbox{\scriptsize tot}} - 1}$
for $1< \Omega_{\hbox{\scriptsize tot}} < 1.1$.
In figure \ref{Fig:Radialfunction}(b) the square of the radial function
$R_{\beta l}^2(\tau)$ is shown for the first eigenvalue
of the binary tetrahedral space $(\beta=7)$,
the binary octahedral space $(\beta=9)$,
and the dodecahedron $(\beta=13)$ for $l=2$.
One observes that the quadrupole suppression due to the first
zero of the radial function is maximal in the case of the
binary tetrahedral space at $\Omega_{\hbox{\scriptsize tot}}\simeq 1.064$,
at $\Omega_{\hbox{\scriptsize tot}}\simeq 1.038$
for the binary octahedral space,
and at $\Omega_{\hbox{\scriptsize tot}}\simeq 1.018$
for the dodecahedral space.
This matches perfectly well to the previously found intervals
on which these models show a strong anisotropy suppression.
Thus, the radial function has an important influence on the
suppression for a given spherical topology.
Note that the zero of the radial function eliminates many eigenfunctions
due to the high degeneracy,
e.\,g.\ the first 13 eigenfunctions in the case of the dodecahedron.
This does not happen in such a dramatic way in models with negative curvature,
where one has no degeneracies at all,
i.\,e.\ all eigenvalues have multiplicity one, in general.
Then the radial function can only suppress a single eigenfunction
and not a ``cluster'' of them.

\begin{figure}[htb]  %  Tetraeder
\begin{center}
%\hspace*{-15pt}
\begin{minipage}{6cm}
\hspace*{-145pt}
\includegraphics[width=8.0cm,angle=270]{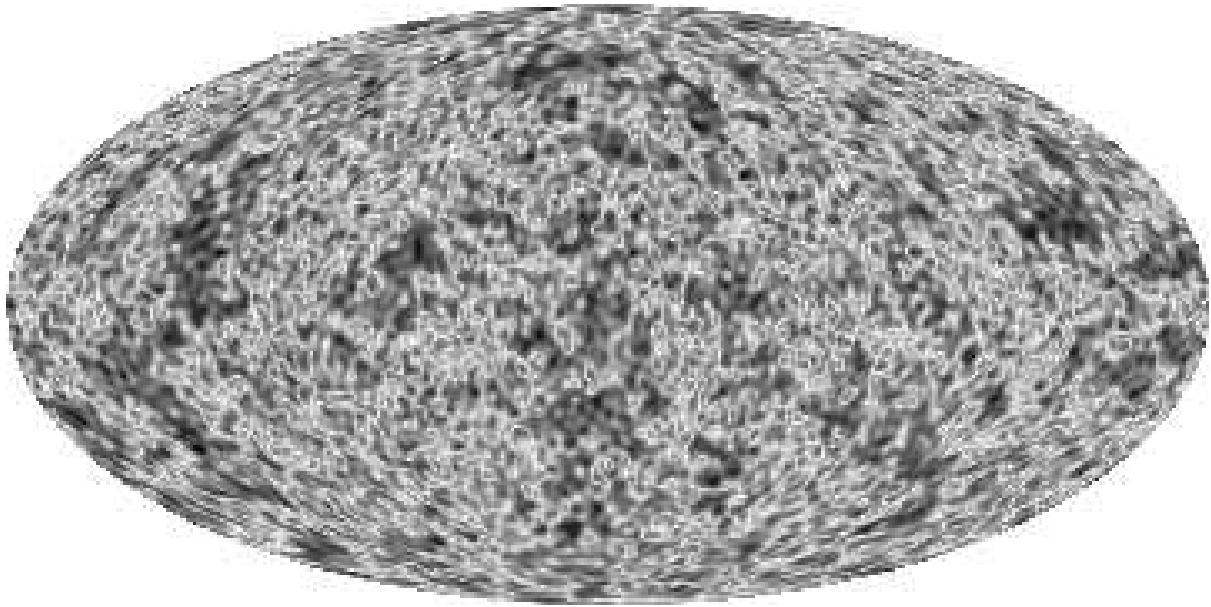}
\end{minipage}
\end{center}
\vspace*{-0.0cm}
\caption{\label{Fig:CMB_Tetrahedron}
The temperature fluctuation $\delta T/T$ of one
realization for the binary tetrahedral group $T^\star$ is shown
($\beta_{\hbox{\scriptsize max}}=155$).
The cosmological parameters
$\Omega_{\hbox{\scriptsize tot}}=1.065$,
$\Omega_\Lambda=0.785$ and $h=70$
are used.
}
\end{figure}

\begin{figure}[htb]  %  Oktaeder
\begin{center}
%\hspace*{-15pt}
\begin{minipage}{6cm}
\hspace*{-145pt}
\includegraphics[width=8.0cm,angle=270]{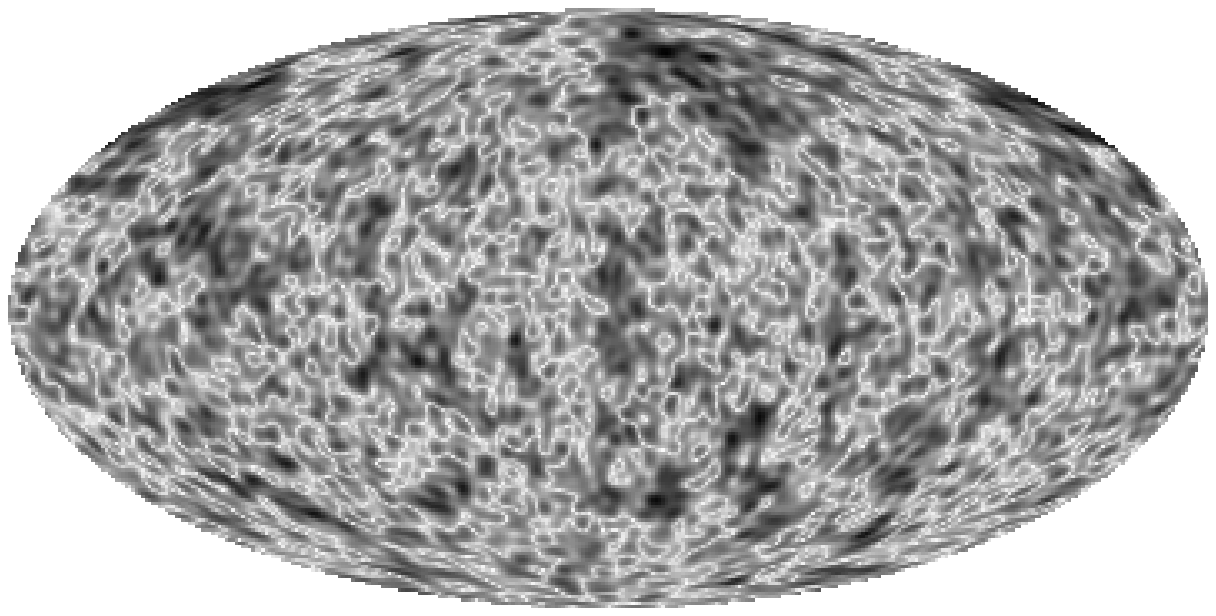}
\end{minipage}
\end{center}
\vspace*{-0.0cm}
\caption{\label{Fig:CMB_Octahedron}
The temperature fluctuation $\delta T/T$ of one
realization for the binary octahedral group $O^\star$ is shown
($\beta_{\hbox{\scriptsize max}}=161$).
The cosmological parameters
$\Omega_{\hbox{\scriptsize tot}}=1.038$,
$\Omega_\Lambda=0.758$ and $h=70$
are used.
}
\end{figure}

\begin{figure}[htb]  %  Dodekaeder
\begin{center}
%\hspace*{-15pt}
\begin{minipage}{6cm}
\hspace*{-145pt}
\includegraphics[width=8.0cm,angle=270]{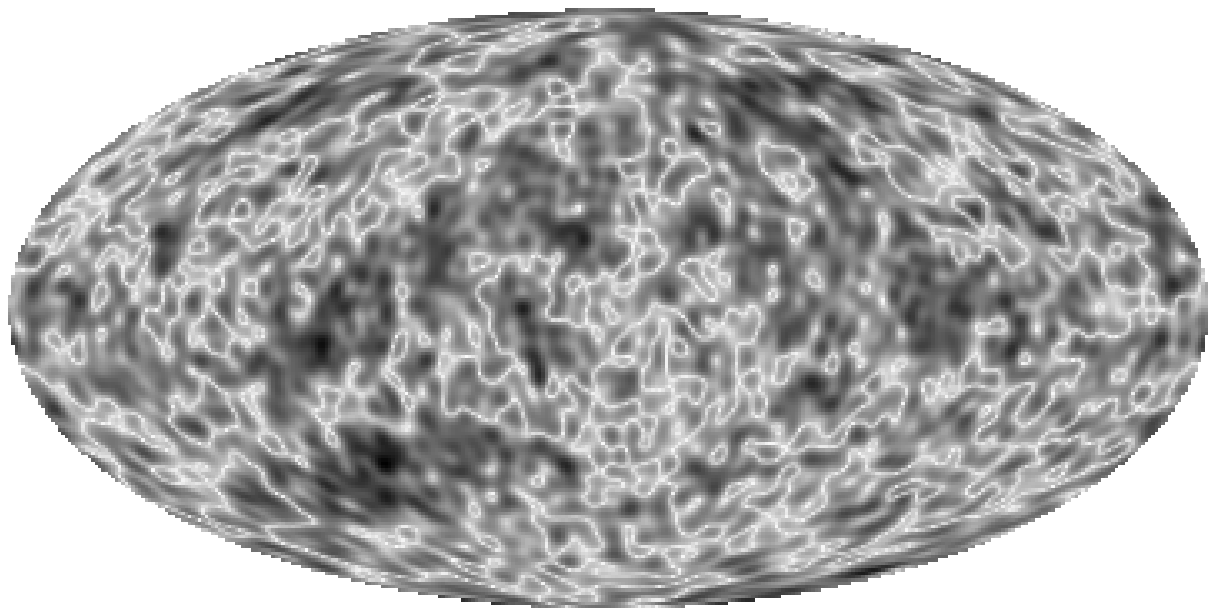}
\end{minipage}
\end{center}
\vspace*{-0.0cm}
\caption{\label{Fig:CMB_Dodecahedron}
The temperature fluctuation $\delta T/T$ of one
realization for the binary icosahedral group $I^\star$ is shown
($\beta_{\hbox{\scriptsize max}}=185$).
The cosmological parameters
$\Omega_{\hbox{\scriptsize tot}}=1.018$,
$\Omega_\Lambda=0.738$ and $h=70$
are used.
}
\end{figure}

The angular power spectrum $\delta T_l^2$ as well as the
$S(\rho)$ statistic lead to the conclusion
that there are three best candidates with respect to spherical spaces,
i.\,e.\ the binary polyhedral spaces
${\cal S}^3/T^\star$, ${\cal S}^3/O^\star$, and ${\cal S}^3/I^\star$.
In figures \ref{Fig:CMB_Tetrahedron} to \ref{Fig:CMB_Dodecahedron}
we show the temperature fluctuation $\delta T/T$ in
the Mollweide projection for these three topological spaces,
where exactly those values of $\Omega_{\hbox{\scriptsize tot}}$ are used
which lead to a strong suppression of large scale power.
In these calculations we have used for the three spaces all modes below
the wave number cut-off $\beta_{\hbox{\scriptsize max}} = 155$,
161 and 185, respectively.

\begin{figure}[htb]  %  Tetraeder
\begin{center}
\vspace*{-90pt}
\begin{minipage}{11cm}
\hspace*{-30pt}\includegraphics[width=11.0cm]{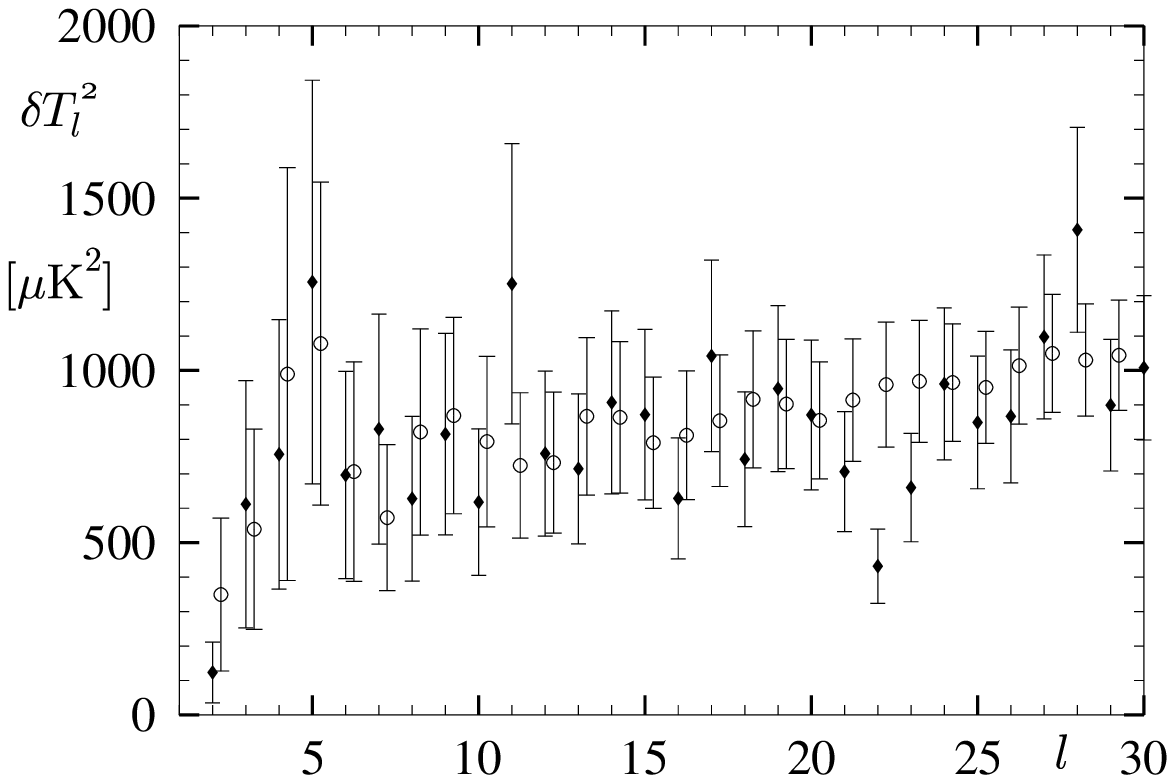}
\end{minipage}
\end{center}
\vspace*{-50pt}
\caption{\label{Fig:Cl_Tetrahedron}
The angular power spectrum $\delta T_l^2$ is shown for the
binary tetrahedral group $T^\star$ (open circles) using the same
cosmological parameters as in figure \ref{Fig:CMB_Tetrahedron}.
The angular power spectrum is shifted by $\Delta l=0.25$ in order
to enable a comparison with the first-year WMAP data (full diamonds).
The $1\sigma$ errors are shown.
}
\end{figure}

\begin{figure}[htb]  %  Oktaeder
\begin{center}
\vspace*{-90pt}
\begin{minipage}{11cm}
\hspace*{-30pt}\includegraphics[width=11.0cm]{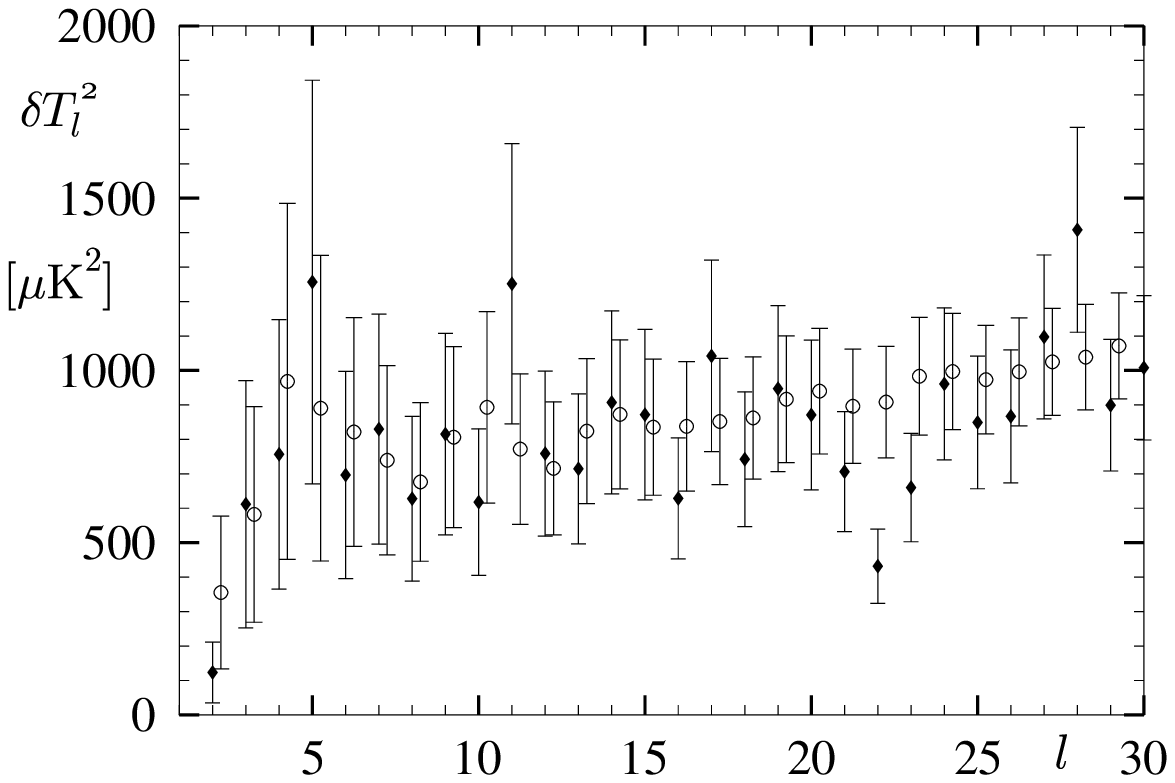}
\end{minipage}
\end{center}
\vspace*{-50pt}
\caption{\label{Fig:Cl_Oktahedron}
The angular power spectrum $\delta T_l^2$ is shown for the
binary octahedral group $O^\star$ (open circles) using the same
cosmological parameters as in figure \ref{Fig:CMB_Octahedron}.
}
\end{figure}

\begin{figure}[htb]  %  Dodekaeder
\begin{center}
\vspace*{-90pt}
\begin{minipage}{11cm}
\hspace*{-30pt}\includegraphics[width=11.0cm]{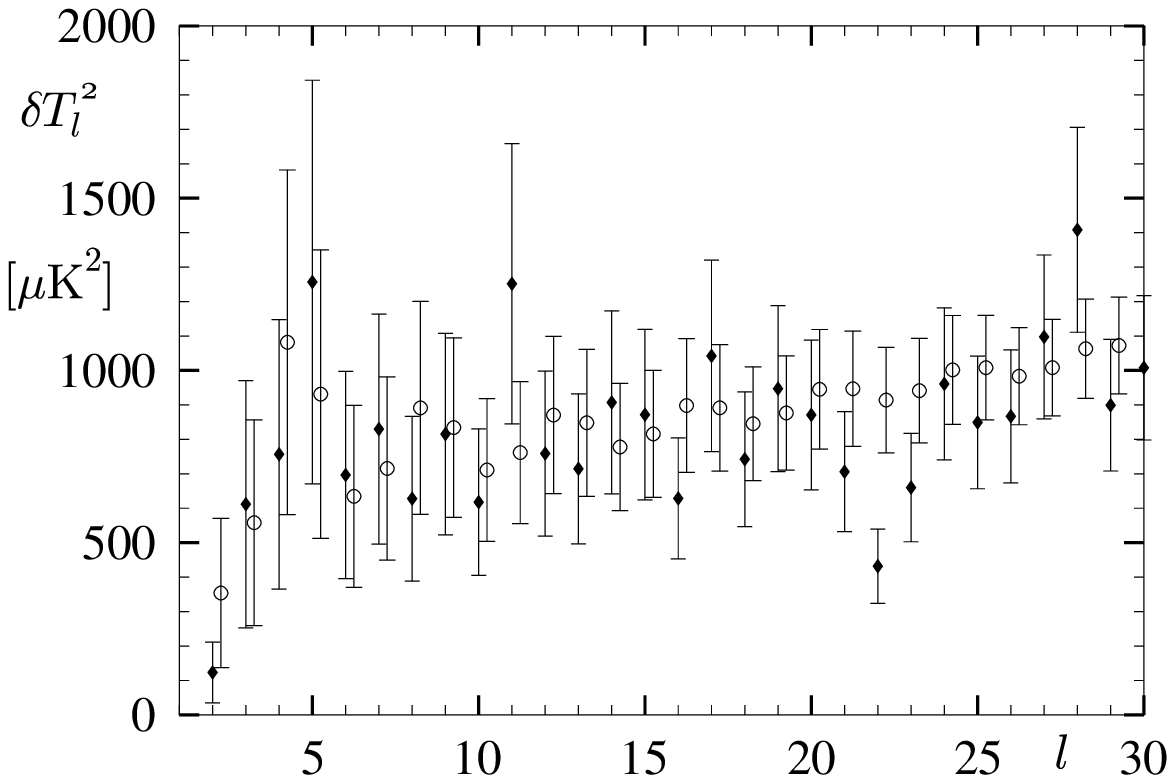}
\end{minipage}
\end{center}
\vspace*{-50pt}
\caption{\label{Fig:Cl_Dodecahedron}
The angular power spectrum $\delta T_l^2$ is shown for the
binary icosahedral group $I^\star$ (open circles) using the same
cosmological parameters as in figure \ref{Fig:CMB_Dodecahedron}.
}
\end{figure}

In figures \ref{Fig:Cl_Tetrahedron} to \ref{Fig:Cl_Dodecahedron}
we show the angular power spectrum $\delta T_l^2$
for the binary polyhedral spaces
${\cal S}^3/T^\star$, ${\cal S}^3/O^\star$, and ${\cal S}^3/I^\star$,
where the same cosmological parameters as in figures
\ref{Fig:CMB_Tetrahedron} to \ref{Fig:CMB_Dodecahedron} are used.
The $1\sigma$ deviations are computed along the lines of
\cite{Gundermann_2005}.
Since the distributions for the lowest multipole moments are asymmetric,
i.\,e.\ not Gaussian,
these error bars should only be considered as providing the order of
magnitude of fluctuations in individual realizations.
In order to facilitate a comparison with the WMAP data,
shown as full diamonds together with their $1\sigma$ errors
not including the cosmic variance,
the spectra of the binary polyhedral spaces are shifted by
$\Delta l=0.25$.
The angular power spectra $\delta T_l^2$ for these three
binary polyhedral spaces are very similar
such that one is faced with a {\it topological degeneracy}
with respect to $\delta T_l^2$.
All three spectra display a good agreement with the WMAP data.

%%%%%%%%%%%%%%%%%%%%%%%%%%%%%%%%%%%%%%%%%%%%%%%%%%%%%%%%%%%%%%%%%%
\section{Conclusion}
%%%%%%%%%%%%%%%%%%%%%%%%%%%%%%%%%%%%%%%%%%%%%%%%%%%%%%%%%%%%%%%%%%
\label{Section_Conclusion}

In this paper we analyse the CMB anisotropy of
homogeneous 3-spaces of constant positive curvature
which are multi-connected and are given by the quotient of
${\cal S}^3$ by a group $\Gamma$ of covering transformations,
i.\,e.\ ${\cal M}={\cal S}^3/\Gamma$.
The motivation is provided by the surprisingly low power
in the CMB anisotropy at the largest scales as measured by
COBE and WMAP and the fact
that the mean value of $\Omega_{\hbox{\scriptsize tot}}$ reported by
WMAP is 1.020 which hints to a positively curved Universe.
In order to explain this low power,
one could modify the primordial power spectrum $P_\Phi(\beta)$,
e.\,g.\ by carefully choosing the inflationary scalar potential,
or by resorting to multi-connected space forms
which give a low CMB anisotropy at the largest scales
due to missing modes compared to the simply-connected ${\cal S}^3$,
in general.

We study all types of homogeneous multi-connected spherical space forms
and find no agreement for the cyclic groups $Z_m$
which show an enhanced power at the largest scales
despite their small volumes.
Also the binary dihedral groups $D^\star_{4m}$ do not lead to models
with a suppression significantly stronger than the simply-connected
${\cal S}^3$ universe.
Thus these models do not seem to provide viable space forms
as a model for our Universe.
This contrasts to the remaining three space forms,
the binary tetrahedral, the binary octahedral as well as
the dodecahedral space forms
which show a sufficiently strong suppression of large scale power
compared to the simply-connected ${\cal S}^3$ universe.
The binary tetrahedral space requires a density
$\Omega_{\hbox{\scriptsize tot}}$
in the range $1.06\dots 1.07$.
Since the WMAP team reported $\Omega_{\hbox{\scriptsize tot}}=1.02\pm0.02$
this model is probably in conflict with the observations.
For the two remaining models the density $\Omega_{\hbox{\scriptsize tot}}$
should be in the range
$\Omega_{\hbox{\scriptsize tot}}=1.03\dots 1.04$
for the binary octahedral space, and
$\Omega_{\hbox{\scriptsize tot}}=1.015\dots 1.02$ for the dodecahedral space.
These values are compatible with the current observations.
Furthermore, we would like to remark that the binary octahedral space
displays a slightly stronger suppression of power than
the dodecahedral space,
as a comparison of figures \ref{Fig:Statistic_Oktaeder} and
figures \ref{Fig:Statistic_Dodekaeder} reveals.

A unique signal for a particular topology is provided by the
so-called circles-in-the-sky-signature proposed in
\cite{Cornish_Spergel_Starkman_1998b}.
Along two circles on the sky which are mapped onto each other
by the group $\Gamma$,
the ordinary Sachs-Wolfe effect produces the same temperature signal.
If there would be no Doppler and integrated Sachs-Wolfe contribution,
see equation (\ref{Eq:Sachs_Wolfe_tight_coupling}),
which disturb this signal,
one would expect a clear sign for a given topology if present.
In \cite{Aurich_Lustig_Steiner_2004c} we study the influence
of the latter two contributions on the circles-in-the-sky-signature
and find that the degradation of the signal is strong enough
such that the topology signal can be swamped.
Therefore, the fact that in \cite{Cornish_Spergel_Starkman_Komatsu_2003} 
no circles are found in the WMAP sky maps,
does not necessarily exclude the binary octahedral or the dodecahedral space
as viable models for our Universe.
In a forthcoming paper we will study the circles-in-the-sky
for the three best multi-connected spherical space forms,
and in particular shall discuss whether a combined circle search
on all circles simultaneously of a given topology can overcome
the degradations.

%%%%%%%%%%%%%%%%%%%%%%%%%%%%%%%%%%%%%%%%%%%%%%%%%%%%%%%%%%%%%%%%%%%%%%%%%%%%
%%%%%%%%%%%%%%%%%%%%%%%%%%%%%%%%%%%%%%%%%%%%%%%%%%%%%%%%%%%%%%%%%%%%%%%%%%%%

\section*{Acknowledgment}

One of us (F.S.) would like to thank the Theoretical Physics Division
of CERN for hospitality.

%%%%%%%%%%%%%%%%%%%%%%%%%%%%%%%%%%%%%%%%%%%%%%%%%%%%%%%%%%%%%%%%%%%%%%%%%%%%
%%%%%%%%%%%%%%%%%%%%%%%%%%%%%%%%%%%%%%%%%%%%%%%%%%%%%%%%%%%%%%%%%%%%%%%%%%%%

\section*{References}

\bibliography{../bib_chaos,../bib_astro}
\bibliographystyle{h-physrev3}

\end{document}